%% file: RefsdalTimeDelays.tex
\newif{\ifchangetext}
\changetextfalse

\documentclass[apj]{emulateapj}
\usepackage{ifthen}
\usepackage{natbib}   
\usepackage{amsmath}  
\usepackage[usenames,dvipsnames]{xcolor}  
\usepackage[colorlinks=true,allcolors=airforceblue]{hyperref}      
\usepackage{longtable}
\usepackage{xspace} 

\definecolor{cerulean}{rgb}{0.0, 0.48, 0.65}
\definecolor{airforceblue}{rgb}{0.36, 0.54, 0.66}
\hypersetup{
colorlinks=true,
allcolors={cerulean}
}

\ifchangetext
  \newcommand{\change}[1]{{ \textcolor{blue}{#1}}}
  \newcommand{\changenote}[1]{{ \textcolor{blue}{#1}}}

\else
  \newcommand{\change}[1]{#1}

  \newcommand{\changenote}[1]{}
\fi

\ifx\pdfoutput\undefined
  \pdffalse
  \DeclareGraphicsExtensions{.eps,.ps}
\else
  \ifnum\pdfoutput=1
    \DeclareGraphicsExtensions{.pdf,.png,.jpg}
  \else
    \DeclareGraphicsExtensions{.eps,.ps}
  \fi
\fi

\defcitealias{Kelly:2015}{K15}
\def\K15{\citetalias{Kelly:2015}\xspace}

\defcitealias{Patel:2014}{P14}
\def\P14{\citetalias{Patel:2014}\xspace}

\def\macs1149{MACS\,J1149.6+2223\xspace}
\def\MACS1149{MACS\,J1149.6+2223\xspace}
\def\pycs{{\tt PyCS}\xspace}

\def\snpy{{\tt SNPy}\xspace}

\def\SNooPy{{\tt SNPy}\xspace}

\def\deg{\ensuremath{^{\circ}}\xspace}

\def\arcsec{\ensuremath{^{\prime\prime}}}

\def\HII{H\,{\sc II}\xspace}

\def\HST{{\it HST}\xspace}

\shorttitle{SN Refsdal Time Delay Measurements}
\shortauthors{Rodney et al.}

\begin{document}

\title{SN Refsdal : Photometry and Time Delay Measurements\\ of the First Einstein Cross Supernova}

\input{Authors}

\input{Abstract}

\input{Introduction}

\input{Observations}

\input{TemplateFitting}

\input{FlexFitting}

\input{Discussion}

\input{Acknowledgments}

\input{RefsdalTimeDelays.bbl}

\input{LongTablePhotometry}

\end{document}

%% file: Authors.tex
\newcommand{\HubbleFellow}{Hubble Fellow}
\newcommand{\Packard}{Packard Fellow}
\newcommand{\CalTech}{California Institute of Technology, 1200 East California Boulevard, Pasadena, CA 91125}
\newcommand{\Cantabria}{IFCA, Instituto de F\'isica de Cantabria (UC-CSIC), Av. de Los Castros s/n, 39005 Santander, Spain}
\newcommand{\IFCA}{\Cantabria}
\newcommand{\JHU}{Department of Physics and Astronomy, The Johns Hopkins University, 3400 N. Charles St., Baltimore, MD 21218, USA}
\newcommand{\Michigan}{Department of Astronomy, University of Michigan, 1085 S. University Avenue, Ann Arbor, MI 48109, USA}
\newcommand{\UCDavis}{University of California Davis, 1 Shields Avenue, Davis, CA 95616}
\newcommand{\UCLA}{Department of Physics and Astronomy, University of California, Los Angeles, CA 90095}
\newcommand{\USC}{Department of Physics and Astronomy, University of South Carolina, 712 Main St., Columbia, SC 29208, USA}
\newcommand{\RCEU}{Research Center for the Early Universe, University of Tokyo, 7-3-1 Hongo, Bunkyo-ku, Tokyo 113-0033, Japan}
\newcommand{\TokyoPhys}{Department of Physics, University of Tokyo, 7-3-1 Hongo, Bunkyo-ku, Tokyo 113-0033, Japan}
\newcommand{\IPMU}{Kavli Institute for the Physics and Mathematics of the Universe (Kavli IPMU, WPI), University of Tokyo, 5-1-5 Kashiwanoha, Kashiwa, Chiba 277-8583, Japan}
\newcommand{\TokyoAstro}{Department of Astronomy, Graduate School of Science, The University of Tokyo, 7-3-1 Hongo, Bunkyo-ku, Tokyo 113-0033, Japan}
\newcommand{\DARK}{Dark Cosmology Centre, Niels Bohr Institute, University of Copenhagen, Juliane Maries Vej 30, DK-2100 Copenhagen, Denmark} 
\newcommand{\INFN}{INFN, Sezione di Bologna, Viale Berti Pichat 6/2, I-40127 Bologna, Italy}
\newcommand{\EHU}{Fisika Teorikoa, Zientzia eta Teknologia Fakultatea, Euskal Herriko Unibertsitatea UPV/EHU}
\newcommand{\Basque}{IKERBASQUE, Basque Foundation for Science, Alameda Urquijo, 36-5 48008 Bilbao, Spain}
\newcommand{\Berkeley}{Department of Astronomy, University of California, Berkeley, CA 94720-3411, USA}
\newcommand{\STScI}{Space Telescope Science Institute, 3700 San Martin Dr., Baltimore, MD 21218, USA}
\newcommand{\Ferrara}{Dipartimento di Fisica e Scienze della Terra, Universit\`{a} degli Studi di Ferrara, via Saragat 1, I-44122, Ferrara, Italy}
\newcommand{\INAF}{INAF, Osservatorio Astronomico di Bologna, via Ranzani 1, I-40127 Bologna, Italy}
\newcommand{\UCSB}{Department of Physics, University of California, Santa Barbara, CA 93106-9530, USA}
\newcommand{\SantaBarbara}{\UCSB}
\newcommand{\Kapteyn}{Kapteyn Astronomical Institute, University of Groningen, Postbus 800, 9700 AV Groningen, the Netherlands}
\newcommand{\WKU}{Department of Physics, Western Kentucky University, Bowling Green, KY 42101, USA}
\newcommand{\IAP}{Institut d’Astrophysique de Paris, UMR7095 CNRS-Universit\'{e} Pierre et Marie Curie, 98bis bd Arago, F-75014 Paris, France}
\newcommand{\ASIAA}{Institute of Astronomy and Astrophysics, Academia Sinica, P.O. Box 23-141, Taipei 10617, Taiwan}
\newcommand{\TokyoKashiwa}{Institute for Cosmic Ray Research, The University of Tokyo, Kashiwa, Chiba 277-8582, Japan}
\newcommand{\Munich}{University Observatory Munich, Scheinerstrasse 1, D-81679 Munich, Germany} 
\newcommand{\KICPStanford}{Kavli Institute for Particle Astrophysics and Cosmology, Stanford University, 452 Lomita Mall, Stanford, CA 94305, USA}
\newcommand{\Andalucia}{Instituto de Astrof\'isica de Andaluc\'ia (CSIC), E-18080 Granada, Spain}
\newcommand{\SaoPaulo}{Instituto de Astronomia, Geof\'isica e Ci\^encias Atmosf\'ericas, Universidade de S\~ao Paulo, Cidade Universit\'aria, 05508-090, S\~ao Paulo, Brazil}
\newcommand{\AMNH}{Department of Astrophysics, American Museum of Natural History, Central Park West and 79th Street, New York, NY 10024, USA}
\newcommand{\NYU}{Center for Cosmology and Particle Physics, New York University, New York, NY 10003, USA}
\newcommand{\Arizona}{Department of Astronomy, University of Arizona, Tucson, AZ 85721, USA}
\newcommand{\Rutgers}{Department of Physics and Astronomy, Rutgers, The State University of New Jersey, Piscataway, NJ 08854, USA}
\newcommand{\NOAO}{National Optical Astronomical Observatory, Tucson, AZ 85719, USA}
\newcommand{\LCOGT}{Las Cumbres Observatory Global Telescope Network, 6740 Cortona Dr., Suite 102, Goleta, California 93117, USA}
\newcommand{\IllinoisAstro}{Astronomy Department, University of Illinois at Urbana-Champaign, 1002 W.\ Green Street, Urbana, IL 61801, USA }
\newcommand{\IllinoisPhysics}{Department of Physics, University of Illinois at Urbana-Champaign, 1110 W.\ Green Street, Urbana, IL 61801, USA }
\newcommand{\AIP}{Leibniz-Institut f\"ur Astrophysik Potsdam (AIP), An der Sternwarte 16, 14482 Potsdam, Germany}

\newcounter{affilct}
\setcounter{affilct}{0}

\makeatletter
\newcommand{\affilref}[1]{%
  \@ifundefined{c@#1}%
    {\newcounter{#1}%
     \setcounter{#1}{\theaffilct}%
     \refstepcounter{affilct}%
     \label{#1}%
     }{}%
  \ref{#1}%
 }
\makeatother

\makeatletter
\newcommand*\affilreftxt[2]{%
  \@ifundefined{c@#1txt}
    {\newcounter{#1txt}%
     \setcounter{#1txt}{1}
     \altaffiltext{\ref{#1}}{#2}
     }{
     }
  }
\makeatother


 \author{S.~A.~Rodney\altaffilmark{\affilref{USC},}}
 \affilreftxt{USC}{\USC}
 \email{srodney@sc.edu}

 \author{L.-G.~Strolger\altaffilmark{\affilref{STScI}}}
 \affilreftxt{STScI}{\STScI}

 \author{P.~L.~Kelly\altaffilmark{\affilref{Berkeley}}}
 \affilreftxt{Berkeley}{\Berkeley}

\author{M.~Brada\v{c}\altaffilmark{\affilref{UCDavis}}}
\affilreftxt{UCDavis}{\UCDavis}

\author{G.~Brammer\altaffilmark{\affilref{STScI}}}
\affilreftxt{STScI}{\STScI}

\author{A.~V.~Filippenko\altaffilmark{\affilref{Berkeley}}}
\affilreftxt{Berkeley}{\Berkeley}

\author{R.~J.~Foley\altaffilmark{\affilref{IllinoisPhysics},\affilref{IllinoisAstro}}}
\affilreftxt{IllinoisPhysics}{\IllinoisPhysics}
\affilreftxt{IllinoisAstro}{\IllinoisAstro}

\author{O.~Graur\altaffilmark{\affilref{NYU},\affilref{AMNH}}}
\affilreftxt{NYU}{\NYU}
\affilreftxt{AMNH}{\AMNH}

\author{J.~Hjorth\altaffilmark{\affilref{DARK}}}
\affilreftxt{DARK}{\DARK}

\author{S.~W.~Jha\altaffilmark{\affilref{Rutgers}}}
\affilreftxt{Rutgers}{\Rutgers}

\author{C.~McCully\altaffilmark{\affilref{LCOGT},\affilref{UCSB}}}
\affilreftxt{LCOGT}{\LCOGT}
\affilreftxt{UCSB}{\UCSB}

\author{A.~Molino\altaffilmark{\affilref{SaoPaulo}},\affilref{Andalucia}}
\affilreftxt{SaoPaulo}{\SaoPaulo}
\affilreftxt{Andalucia}{\Andalucia}

\author{A.~G.~Riess\altaffilmark{\affilref{JHU},\affilref{STScI}}}
\affilreftxt{JHU}{\JHU}
\affilreftxt{STScI}{\STScI}

\author{K.~B.~Schmidt\altaffilmark{\affilref{UCSB},\affilref{AIP}}}
\affilreftxt{UCSB}{\UCSB}
\affilreftxt{AIP}{\AIP}

\author{J.~Selsing\altaffilmark{\affilref{DARK}}}
\affilreftxt{DARK}{\DARK}

\author{K.~Sharon\altaffilmark{\affilref{Michigan}}}
\affilreftxt{Michigan}{\Michigan}

\author{T.~Treu\altaffilmark{\affilref{UCLA},\affilref{Packard}}}
\affilreftxt{UCLA}{\UCLA}
\affilreftxt{Packard}{\Packard}

\author{B.~J.~Weiner\altaffilmark{\affilref{Arizona}}}
\affilreftxt{Arizona}{\Arizona}

\author{A.~Zitrin\altaffilmark{\affilref{CalTech},\affilref{HubbleFellow}}}
\affilreftxt{CalTech}{\CalTech}
\affilreftxt{HubbleFellow}{\HubbleFellow}

%% file: Abstract.tex
\begin{abstract}

We present the first year of Hubble Space Telescope imaging of
the unique supernova (SN) 'Refsdal', a gravitationally lensed SN at
$z=1.488\pm0.001$ with multiple images behind the galaxy cluster
\macs1149.  The first four observed images of SN Refsdal
  (images S1--S4) exhibited a slow rise (over $\sim150$ days) to
reach a broad peak brightness around 20 April, 2015.
Using a set of light curve templates constructed from  SN
1987A-like peculiar Type II SNe, we measure time delays for the four
images relative to S1 of 4$\pm$4 (for S2), 2$\pm$5 (S3), and
  24$\pm$7 days (S4). The measured magnification ratios relative to S1
  are 1.15$\pm$0.05 (S2), 1.01$\pm$0.04 (S3), and 0.34$\pm$0.02 (S4).
None of the template light curves fully
captures the photometric behavior of SN Refsdal, so we also derive
complementary measurements for these parameters using
  polynomials to represent the intrinsic light curve shape. These more
  flexible fits deliver fully consistent time delays of 7$\pm$2 (S2), 0.6$\pm$3  (S3), and 27$\pm$8 days (S4).  The lensing
  magnification ratios are similarly consistent, measured as
  1.17$\pm$0.02 (S2), 1.00$\pm$0.01 (S3), and 0.38$\pm$0.02 (S4). We
compare these measurements against published predictions from lens
models, and find that the majority of model predictions are
  in very good agreement with our measurements.   Finally, we discuss
avenues for future improvement of time delay measurements -- both for
SN Refsdal and for other strongly lensed SNe yet to come.

\end{abstract}

%% file: Introduction.tex
\section{Introduction}

\begin{figure*} 
   \centering
   \includegraphics[width=\linewidth]{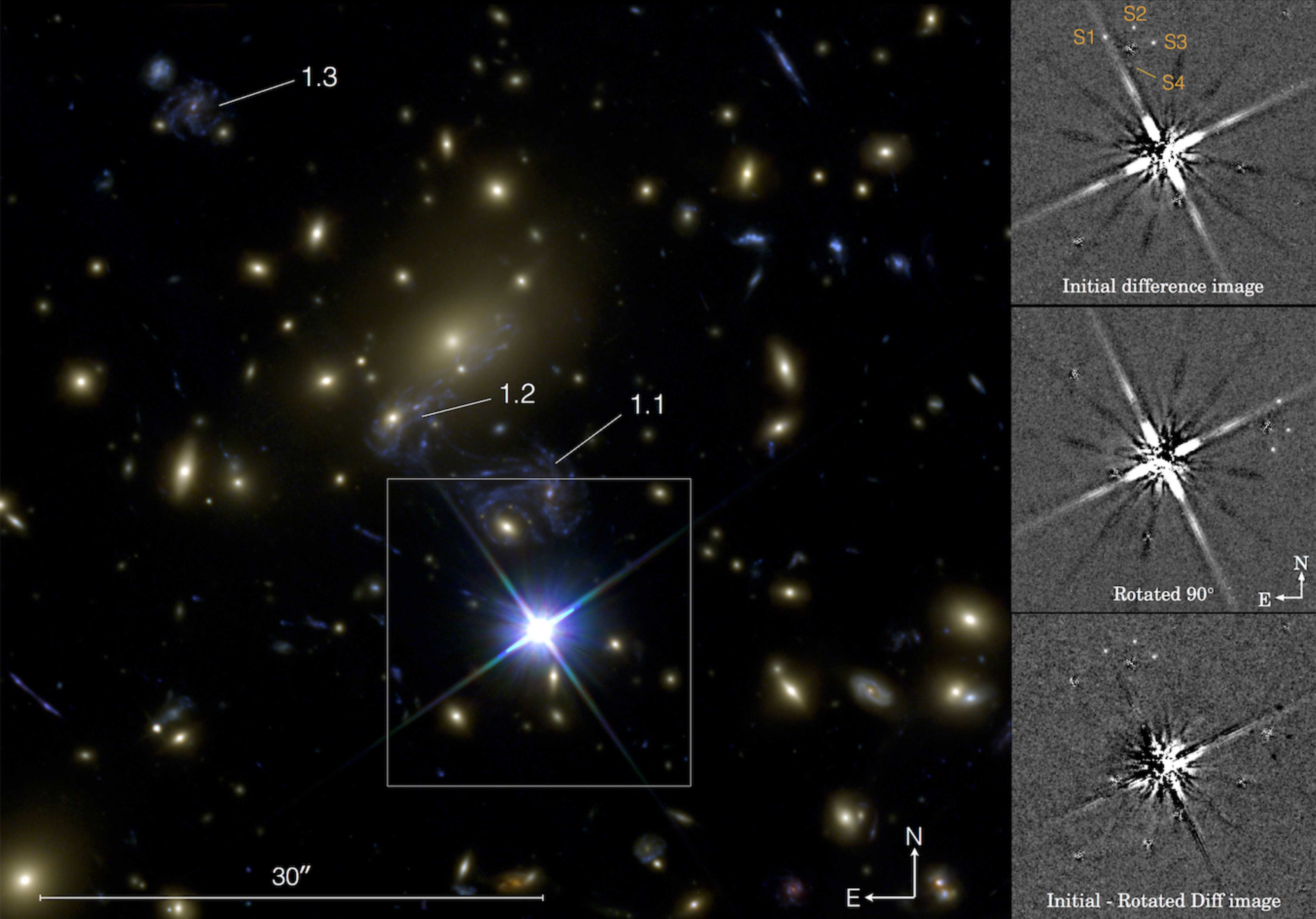}
   \caption{The \macs1149 field, showing the positions of the three
     primary images of the SN Refsdal host galaxy (labeled
     1.1, 1.2 and 1.3).  SN Refsdal appears as four point sources in
     an Einstein Cross configuration in the southeast spiral arm of
     image 1.1.  The highlighted box is shown at the same scale in
     panels on the right side, which illustrate the removal of
     contaminating diffraction spikes from a difference image. Each
     difference image is centered on the location of the contaminating
     star (top panel), then rotated clockwise by 90\deg (middle
     panel). The rotated difference image is then subtracted from the
     initial difference image, removing most of the flux from the
     contaminating diffraction spike at the location of the SN Refsdal
     point sources (bottom panel).
   \label{fig:macs1149field}}
\end{figure*}

The discovery of SN Refsdal, the first strongly-lensed supernova (SN)
resolved into multiple images, was described by
\cite{Kelly:2015a}. SN~Refsdal was located in the arm of a face-on
spiral host galaxy at $z=1.49$.  This spiral arm is distorted into an
Einstein ring by the gravitational potential of a foreground
elliptical galaxy.  That elliptical galaxy lens also resides within
\macs1149, a strong-lensing galaxy cluster at $z=0.54$ that is fast
becoming one of the crown jewels of the Massive Cluster Survey
~\citep{Ebeling:2001,Ebeling:2007}.  The galactic-scale lens,
augmented by the cluster lens, causes SN~Refsdal to appear to us as
four images with separations of $\sim$2\arcsec, arranged in an
``Einstein Cross'' configuration (see Figure~\ref{fig:macs1149field})
reminiscent of the quadruply-imaged quasar that originated this term
\citep{Huchra:1985,Adam:1989}.

The host galaxy of SN~Refsdal, is
itself strongly lensed by the \macs1149 cluster, and was identified as
a particularly spectacular example of a multiply-imaged galaxy in some
of the earliest lens modeling efforts \citep{Smith:2009,Zitrin:2009b}.
Due to the spatial magnification afforded by the cluster lens, this
galaxy has provided a rare opportunity to study the substructure of a
$z=1.5$ galaxy at scales down to $\sim$100 pc.  \change{This galaxy} shows
evidence for active star formation
\citep{Smith:2009,Livermore:2012,Livermore:2015} with a young stellar
population at the SN Refsdal position \citep{Adamo:2013}.
\citet{Yuan:2011} reported a steep metallicity gradient for the
galaxy, and \citet{Yuan:2015} measured a low metallicity from nine
\HII regions at similar galactocentric radii.  Using integral field
spectroscopy with the VLT MUSE spectrograph, \citet{Karman:2015} found
\ion{Mg}{2} emission at the SN Refsdal position, and inferred from the
    [\ion{O}{2}] to \ion{Mg}{2} ratio that the SN exploded in a low
    metallicity and high ionization environment.

This host galaxy presents at least three distinct images in the plane
of the \macs1149 field, and the image in which SN Refsdal was
discovered is typically labeled as image 1.1 \citep{Smith:2009}.  Lens
models consistently indicate that the second image of this galaxy,
1.2, is a trailing image
\citep[e.g.][]{Kelly:2015a,Oguri:2015,Sharon:2015,Treu:2015b}, and
indeed a new transient source appeared at the expected location in
December 2015.  This new source is consistent with being the predicted
reappearance image, SX \citep{Kelly:2015d}.  A third image of the host
galaxy, image 1.3, is understood to be a leading image, and the first
image of SN Refsdal most likely appeared there some 20$+$ years ago,
\change{although the available archival \HST\ observations cannot
  confirm or refute this expectation.}

Since the discovery of SN Refsdal, many lens modeling teams
  have produced updated lens models and generated predictions for the
  SN time delays and magnifications, in some cases taking advantage of
  the very deep imaging and spectrosopic data from the Hubble Space Telescope (\HST) and other
  observatories
  \citep{Diego:2015b,Grillo:2015b,Jauzac:2015c,Kawamata:2015,Oguri:2015,Sharon:2015}.
  \citet{Treu:2015b} describes the collaborative development of
  some of these updated lens models by five independent
    teams, and highlights the rare opportunity for a true blind test
  of these models. By generating these predictions in advance of the
  reappearance of SN Refsdal as image SX, the modelers have provided
    falsifiable predictions that can be directly confronted with a
    true measurement of the time delays and magnification ratios.  An
    initial comparison based on the first detection of the
    reappearance image SX showed that several of the models are 
    consistent with observations \citep{Kelly:2015d}. However, a
    complete evaluation will need to await the full light curve of
    image SX, which will be collected over the coming year with an
    ongoing \HST imaging campaign (GO-PID:14199, PI:P.\,Kelly).

The comparison of SN Refsdal observations against model predictions is
similar in concept to \change{previous} tests of lens models using Type Ia SNe
as standardizable candles \citep{Patel:2014,Nordin:2014,Rodney:2015a}.
For this small sample of lensed Type Ia SNe, it was possible to
constrain the absolute magnification along a single sight line and
compare to model predictions.  SN Refsdal is not a Type Ia SN
\citep{Kelly:2015a}, but instead appears to be a peculiar Type II SN
\citep{Kelly:2015c} so the absolute magnification can not be
determined to the same level of precision.  However, with
magnification factors as high as $\mu$~20, SN Refsdal is much more
strongly lensed than any cluster-lensed SNe previously
seen. Furthermore, as the only known SN with resolved multiple images,
SN Refsdal offers the first chance to test lens models using time delay
measurements. This exercise will inform future prospects for using
strongly lensed SNe as probes of both galaxy and cluster lenses, and
may be valuable for understanding the prevalence of microlensing
effects \citep[][and see
  Section~\ref{ssec:Microlensing}]{Dobler:2006}.

In this paper we present the first year of \HST photometry of the
first four observed images of SN~Refsdal. A companion paper
  \citep{Kelly:2015c} describes the classification of SN Refsdal as a
peculiar Type II SN similar to SN 1987A, based on the \HST light curve as well as \HST and VLT
  spectroscopy.  An outline of the content of this paper is as
follows: Section~\ref{sec:Observations} describes the \HST imaging
observations, data processing, and photometry.  To measure the
gravitational lensing time delays and magnification ratios, in
Section~\ref{sec:LightCurveTemplateFitting} we use light curve
templates, and in Section~\ref{sec:FlexFitting} we use flexible
polynomial light curve models. Finally, we offer a summary and
discussion of results in Section~\ref{sec:Discussion}.

%% file: Observations.tex
\section{Observations and Photometry}\label{sec:Observations}

The imaging observations of SN~Refsdal presented here were all
obtained with \HST using the Wide-Field Camera 3 (WFC3) with the
infrared (IR) and UV-optical (UVIS) detectors, and the Advanced Camera
for Surveys (ACS) Wide Field Camera (WFC).  Here we present all
\HST\ observations from the discovery epoch on 10 November 2014
through the observations of 15 Nov 2015, one year later.  The
\macs1149 field was continuously observed by \HST throughout this
period with a span of no more than 3 weeks between each visit, except
during the period from 21 July 2015 to 30 October 2015, when the field
was too close to the Sun for safe observations with \HST.

As detailed in \citet{Kelly:2015a}, SN~Refsdal was discovered in
images collected for the Grism Lens-Amplified Survey from Space
(GLASS) program
\citep{Schmidt:2014,Treu:2015a}.\footnote{\url{http://glass.astro.ucla.edu}
  and \url{https://archive.stsci.edu/prepds/glass}} The \macs1149
cluster field was subsequently and extensively observed in the course
of the \HST Frontier Fields program (HFF, GO-13504; PI:
Lotz)\footnote{\url{http://www.stsci.edu/hst/campaigns/frontier-fields}} providing
a very rich set of optical and near-IR imaging. The HFF imaging
cadence was supplemented by observations from the Frontier Fields
Supernova program (FrontierSN, GO-13790; PI: Rodney), which extended
the WFC3-IR imaging beyond the end of the HFF campaign to complete the
near-IR light curves at later times.  Additional imaging -- as well as
deep (34 orbits) grism observations -- was provided by an \HST
follow-up program allocated through Director's Discretionary time
\citep[GO/DD-14041; PI: Kelly;][]{Kelly:2015c}.  The \HST monitoring
of this field continues under an ongoing imaging program (GO-14199;
PI: Kelly).

To construct multi-color light curves of the four SN Refsdal sources,
we first sorted the available observations into 45 imaging epochs,
each of which contains observations that were collected within 2
observer-frame days of each other.  We then processed the \HST image
data using tools from the {\tt DrizzlePac} software
suite.\footnote{\url{http://drizzlepac.stsci.edu}} The same-filter
observations for each epoch were registered to a common
astrometric frame using {\tt TweakReg} and combined with {\tt
  AstroDrizzle} \citep{Fruchter:2010}.  The composite images were
drizzled to a pixel scale of 0\farcs06/pixel for WFC3/IR and
0\farcs03/pixel for WFC3/UVIS and ACS/WFC data.  Most of the resulting
composite images in IR bands have total effective exposure times of
$\sim$1200 sec (half of the exposure time typically available in one \HST
orbit).  The ACS-WFC observations are primarily from the deeper HFF
visits, and have composite exposure times of $\sim$5000 sec (two full
\HST orbits).  Table~\ref{tab:Photometry} lists all the composite
observation epochs, including exposure times.

As the final step in the data processing pipeline, we subtracted off a
template image to remove contaminating light from the static
foreground cluster galaxies and SN Refsdal's host galaxy. These
templates were constructed from \HST images collected prior to 15
April 2014, with contributions from the GLASS and HFF programs, but
primarily from data collected as part of the Cluster Lensing And
Supernova survey with Hubble (CLASH, GO-12068; PI Postman,
\citealt{Postman:2012}).  As can be seen in
Figure~\ref{fig:macs1149field}, the location of SN refsdal is
uncomfortably close to a 15th magnitude star \change{USNO
  1050-06589751 (R.A., Decl = 11:49:35.41, 22:23:38.0, \citealt{Monet:2003})}
In some \HST imaging visits the
telescope was oriented such that diffraction spikes from this star
overlapped the position of one or more of the SN Refsdal source
positions.  In the template images, this impacted only image S4 in the
F125W filter.  To resolve this we generated a special set of templates
that excluded those spike-contaminated template images. These slightly
shallower templates were used only to gather photometry for image S4.

For epochs with SN Refsdal present, when the telescope orientation led
to diffraction spike contamination of one or more of the four images,
we cannot simply discard the contaminated observations.  Instead,
these images were processed through an additional ``despiking''
procedure to enable less biased photometric measurements.  The
diffraction spike pattern on \HST in the WFC3-IR detector is close to
symmetric about both axes, so we could generate a rough model for the
contaminating spike by centering the image on the star, and then
rotating the difference image by 90\deg\ in a clockwise direction.  We
then remove the spike by subtracting the rotated difference image from
the original unrotated version, which effectively removes the majority
of the contaminating flux at the Refsdal source locations, as shown in
Figure~\ref{fig:macs1149field}.  We examined modifications to this
approach, such as using a 180\deg\ or 270\deg\ rotation, or a median
of three rotated versions.  We found that a single 90\deg\ clockwise
rotation was most effective, and alternatives did not substantially
affect the resulting photometry.  \change{By inserting and recovering artificial point sources} in
  the spike-contaminated regions, we have confirmed that this
  despiking procedure does increase the statistical uncertainty of our
  photometric measurements, but results in a net improvement by
  reducing the potential for systematic biases.

\subsection{Photometry}\label{sec:Photometry}

For our photometric measurements on the difference images, we used the
{\tt PythonPhot}\footnote{\url{https://github.com/djones1040/PythonPhot}}
software package \citep{Jones:2015}, developed in part for use on
other high-$z$ SNe observed with \HST \citep[e.g.][]{Rodney:2015a,
  Rodney:2015b}. We measured the flux using a 
point spread function (PSF) fitting procedure similar to DAOPHOT~\citep{Stetson:1987}.  As
in \citet{Rodney:2015a,Rodney:2015b}, we used an empirical
PSF model generated from the \HST imaging of the G2V standard star
P330E, observed in a series of \HST calibration programs.

\begin{figure*}
   \centering
   \includegraphics[width=\textwidth]{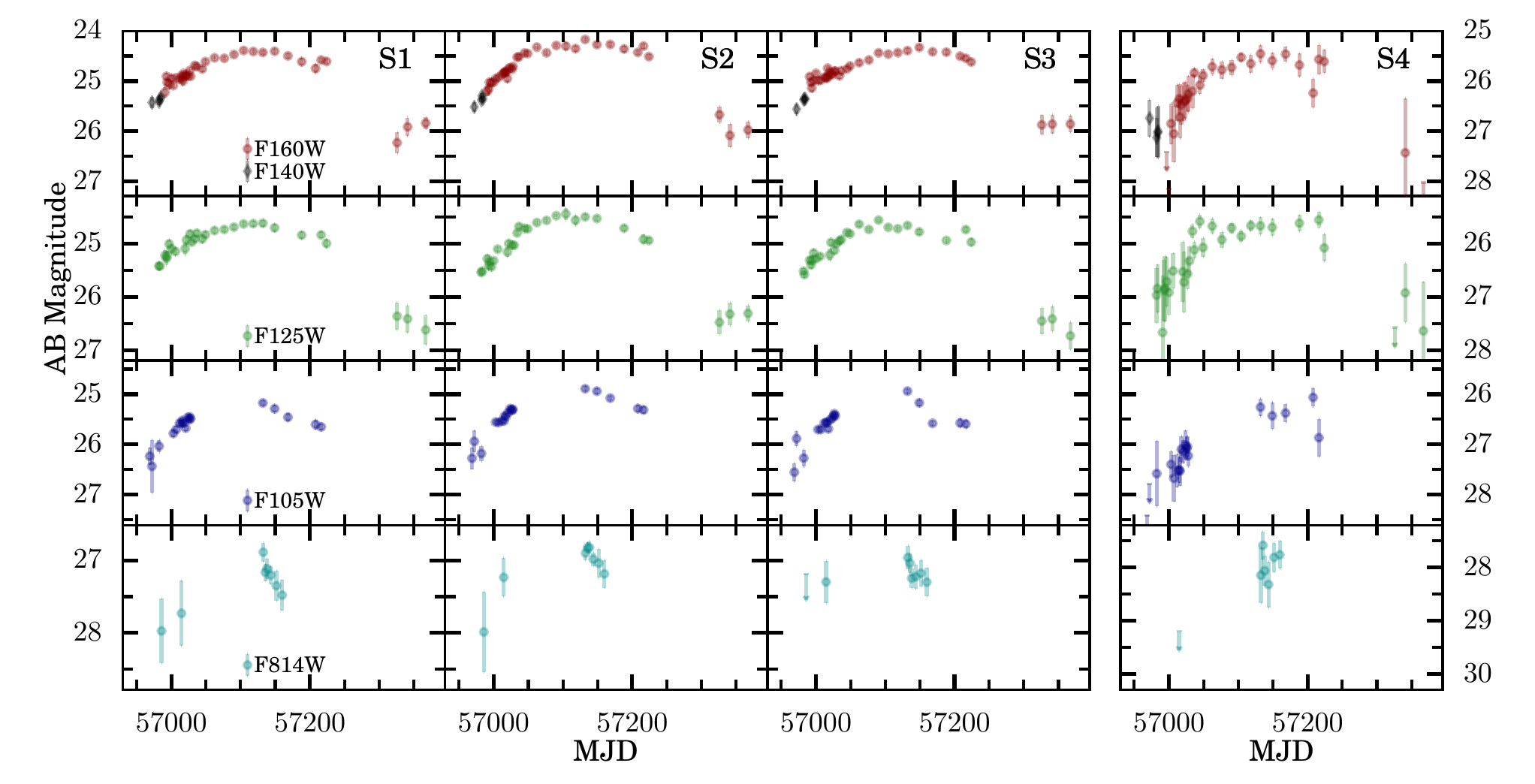}
   \caption{The observed light curves of SN Refsdal images S1--S4. Each panel shows AB magnitudes plotted against observer-frame days. Each column shows the light curve of one of the SN Refsdal images, S1-S4 from left to right, with F160W and F140W in the top row, then F125W, F105W, and F814W separately in rows two to four, respectively.
   \label{fig:LightCurves}}
\end{figure*}

To estimate photometric uncertainties in each image, we planted and
extracted 500 fake stars (copies of the model PSF) at random locations
in the region defined by the sky annulus.  We measured the flux
density of each fake star with PSF fitting and fit a normal
distribution to the histogram of recovered fake star flux densities.
We then define two components of the uncertainty from the best-fit
normal distribution. First, $\delta f_{\mu}$ is the difference between
the measured mean of the distribution and the value of the flux
assigned to all the planted fake stars, which is typically very close
to zero but can give an estimate of systematic biases in cases where
the sky region around the SN is strongly contaminated by diffraction
spikes or residuals from the lensing galaxy.  Second, $\delta
f_{\sigma}$ is the standard deviation of the best-fit normal
distribution, and gives an empirical measure of the uncertainty due to
sky noise and detector read noise. A final uncertainty component is
$\delta f_{\nu}$, the Poisson noise error, computed from the total
count of photons measured in the PSF fit or the aperture.  These are
added in quadrature to give the total uncertainty, where $\delta f^2$
= ${ \delta f_{\mu}^2 + \delta f_{\sigma}^2 + \delta f_{\nu}^2}$.

These photometric measurements are reported in
Table~\ref{tab:Photometry} and Figure~\ref{fig:LightCurves} shows the
resulting multi-band light curves for images S1-S4.  In
Table~\ref{tab:Photometry} we mark with an asterisk any photometric
measurement that was collected from despiked images prepared as in
Figure~\ref{fig:macs1149field}.

\subsection{Photometric Classification}\label{ssec:PhotometricClassification}

  Details of the classification of SN Refsdal are presented in a
  companion paper \citep{Kelly:2015c}, which draws on all available
  photometric and spectroscopic data.  There we present
    spectroscopy of SN~Refsdal from \HST and the Very Large Telescope
    (VLT), taken roughly 75 days apart in the rest-frame. In all
    spectra, we identify broad H$\alpha$ emission consistent with a
    Type II SN at the redshift of the host galaxy.  This
    classification is reinforced by the slow rise to peak brightness
    (over $\sim$150 days) observed in all four SN Refsdal light
    curves \change{(see Figure~\ref{fig:LightCurves}).}
      This light curve shape is most consistent with the
    well-studied archetype, SN 1987A, a peculiar Type II SN that is
    understood to be the explosion of a blue supergiant star.
    Although no single line of evidence provides a definitive
    classification of the SN sub-type, the preponderance of evidence
    indicates that SN Refsdal is a Type II SN, and most likely a
    member of the rare SN 1987A-like sub-class.

%% file: TemplateFitting.tex
\section{Light Curve Template Fitting}
\label{sec:LightCurveTemplateFitting}

 In recent years, high precision time delays have been
  measured for a growing sample of multiply-imaged quasars, using
  increasingly sophisticated observations and techniques
  \citep[e.g.,][]{Fassnacht:2002,Kochanek:2006,Courbin:2011,Eulaers:2013,Tewes:2013b}
  Measuring time delays from lensed SNe like SN Refsdal should in principle be
  much simpler than is typically the case for lensed quasars. 
  The time variation of quasars is stochastic, being driven by
  essentially random events on the accretion disk of the central
  supermassive black hole, so the intrinsic shape of a quasar light
  curve can not be known {\it a priori}.  As such, quasar time delay
  methods must adopt a very flexible function to describe the light
  curve, and rely on purely empirical constraints
  \citep[e.g.,][]{Tewes:2013a,Liao:2015}. In contrast, for lensed SNe
  it should typically be possible to classify the SN based on both
  photometry and spectroscopy, and then identify a well-matched SN
  light curve template.  In that case, a template-based approach for
  time delay measurements will almost always be preferable to using a
  flexible function, as the template provides a strong informative
  prior for the intrinsic light curve shape.

In this section we derive our first measurements of relative
  time delays and magnification ratios from the SN Refsdal data using
  light curve template matching.  This approach makes the assumption
that the SN Refsdal light curve shape can be well approximated by a
light curve model based on well-studied SNe from the nearby
Universe. A second set of time delay and magnification measurements
using different assumptions will be presented in
Section~\ref{sec:FlexFitting}.

\subsection{Template Fitting Method}\label{ssec:TemplateFittingMethod}

As described in Section~\ref{ssec:PhotometricClassification},
SN~Refsdal's slow rise to maximum light is clearly inconsistent with
the rise times for the most common SN types~\citep[e.g., Ia, Ib/c,
  II-P, and II-L; cf.][]{Li:2011a}.  For completeness, we also
evaluated the quality of fit from these normal SN classes, using a
library of 42 templates drawn from the Supernova Analysis software
suite \citep[SNANA][]{Kessler:2009a}. Unsurprisingly, the photometric
peculiarity of SN Refsdal is born out quantitatively, as our light
curve models for these normal SN sub-classes are highly incompatible
with the data, returning a $\chi^2$ per degree of freedom $\nu$
$\chi^2_{\nu}\gg50$.  These models are therefore formally rejected,
and the remainder of our analysis focuses on the peculiar SN
1987A-like sub-class, which provides the best matches to the observed
shape of the SN Refsdal light curve.

We constructed templates based on the prototype SN 1987A itself
\citep{Hamuy:1990}, and also using the 87A-like events SN 1998A
\citep{Woodings:1998,Pastorello:2005}, SN 2000cb
\citep{Hamuy:2001,Kleiser:2011}, SN 2006V and SN 2006au
\citep{Taddia:2012}, and SN 2009E \citep{Pastorello:2012}.  All of
these template SNe have well-sampled light curve coverage in the $B$,
$V$ and $R$ bands extending over at least 80 days in the rest-frame.
\change{As detailed below,} each template was corrected to appear as
it would at the redshift of SN Refsdal and through the observed \HST
passbands. We then implemented a Bayesian parameter estimation
framework \citep[similar to][]{Rodney:2009,Rodney:2010a} to
simultaneously find the color corrections needed to match each model
to the SN Refsdal data, as well as the best-fit time delays and
magnifications for all four SN Refsdal images.

The model light curves are defined using:
\begin{equation}
m(\lambda',t')=M(\lambda,t)+K(\lambda,t;\lambda')+C_{\lambda},
\end{equation}
\noindent where the time $t$ is the rest-frame age relative to the
date of peak brightness in the rest-frame $R$ band, MJD$_{\rm pk}$ (a
free parameter in the model). The rest-frame time is dilated to the
observer frame using $t'=t\,(1+z)$.  The model apparent magnitude in
an observed passband at given observed age, $m(\lambda',t')$, is
governed by a model absolute magnitude in a model passband at the
model's rest-frame age, $M(\lambda,t)$, corrected to an observed
passband with $K(\lambda,t;\lambda')$ \citep[see][for an example of
  the applied k-correction]{Strolger:2015}. A magnitude shift
$C_{\lambda}$ is then added as a separate free parameter for each
photometric passband, which  accounts for both cosmological
dimming and any color difference between the model and SN Refsdal (due
to dust extinction or intrinsic color differences).  Linear
  interpolation is used to infer model magnitudes between observed
  points in the template light curves.

To take into account the gravitational lensing effects, we
include six more free parameters that are applied as corrections to
the observed data: three time shifts $\Delta t_{i}$ and three
achromatic magnitude shifts $\Delta m_{i}$ that give the time delays and
magnifications of the three sources $i=$(S2, S3, S4) relative to our
reference source S1. The model light curves are then
  simultaneously compared to all four SN Refsdal sources in the
F105W, F125W and F160W bands (rest-frame $B$, $V$ and $R$) to derive a
likelihood distribution from each light curve template  T$_k$, using

\begin{equation}
p({\rm\bf{D}}|{\rm T_k,{\bf \theta}}) = \prod_i \frac{p({\bf \theta})}{\sqrt{2\pi}\sigma_i} e^{-(m_{\rm obs}(t_i) - m_{k}(\lambda_i,t_i))^2 / (2\sigma_i^{2})}.
\label{eqn:Likelihood}
\end{equation}

\noindent Here {\bf$\theta$} denotes the set of 10 free
  parameters: date of peak brightness, three relative time delays
  $\Delta$t, three lensing magnitude shifts $\Delta m_{i}$, and three
  ``color'' shifts $C_{\lambda}$ for the three photometric bands used.  We
  use flat priors $p({\bf \theta})$ for all of the parameters, with
  time shifts allowed over the range [-100,100], and magnitude shifts
  in the range [-3,3].  The product is over all observed epochs $t_i$,
  and the uncertainty for each epoch $\sigma_i^2$ is a quadratic sum
  of the photometric uncertainty and a ``model uncertainty'' of 0.15
  mag.  This term accounts for the fact that there is no perfect light
  curve analog available within our limited template library, due to
  the diversity and rarity of SN 1987A-like events
  \citep[e.g.,][]{Pastorello:2012,Taddia:2012}.  The choice of 0.15
  mag in all epochs follows \citep{Rodney:2009}, where that value was
  found to approximately compensate for a similarly sparse library of
  core collapse light curve templates.  Including this model
  uncertainty removes our ability to independently test for goodness
  of fit, so in this section we are making the strong assumption that
  the templates, blurred by this error term, are a good model for the
  observations.

Alternatively, we can set the model uncertainty to zero, which
effectively assumes that all possible SN 1987A-like light curve shapes
are represented within our set of six viable templates.  In this case,
all the best-fitting models return $\chi^2_\nu\gtrsim7$, indicating
that the models are poor representations of the intrinsic SN light
curve shape.  Nevertheless, with no model uncertainty term we still
find that the range of time delay and magnification estimates are
consistent with the values derived using 0.15 mag for the model
uncertainty.

To sample the likelihood distributions defined by
Equation~\ref{eqn:Likelihood} over the ten-dimensional parameter
space, we use the Markov Chain Monte Carlo ensemble sampling tools
from the {\tt emcee} software package \citep{Foreman-Mackey:2013}.

\subsection{Template Fitting Results}\label{ssec:TemplateFittingResults}

A summary of the template fitting results is given in
Table~\ref{tab:TemplateFitResults}.  To derive a single set
  of measurements from these models, we use the approach of Bayesian
  Model Averaging
  \citep[BMA;][]{Leamer:1978,Raftery:1995,Draper:1995}, which provides a weighted average for each parameter of interest, incorporating the posterior probabilities in the weighting.  The BMA
  posterior mean and variance for each parameter $\phi$ in {\bf
    $\theta$} are given by

  \begin{multline}
E[\phi|{\bf D}] = \sum_k \hat{\phi}_k p(T_k|{\bf D})\\
{\rm Var}[\phi|{\bf D}] = \sum_k ({\rm Var}[\phi|{\bf D},T_k] + \hat{\phi}_k^2)p(T_k|{\bf D}) - E[\phi|{\bf D}]^2
\label{eqn:BMA}
  \end{multline}

\noindent where $\hat{\phi}_k=E[\phi|D,T_k]$ is the
  expectation value assuming template $T_k$ is the correct model.  The
  posterior probability values $p(T_k|{\bf D})$ (reported in column two
  of Table~\ref{tab:TemplateFitResults}) are computed by applying
  Bayes' Theorem with a flat prior $p(T_k)$ for all templates
  \citep[see, e.g.,][for further discussion of the BMA
    method]{Hoeting:1999}.

\begin{figure*}[pht] 
   \centering
   \includegraphics[width=\textwidth]{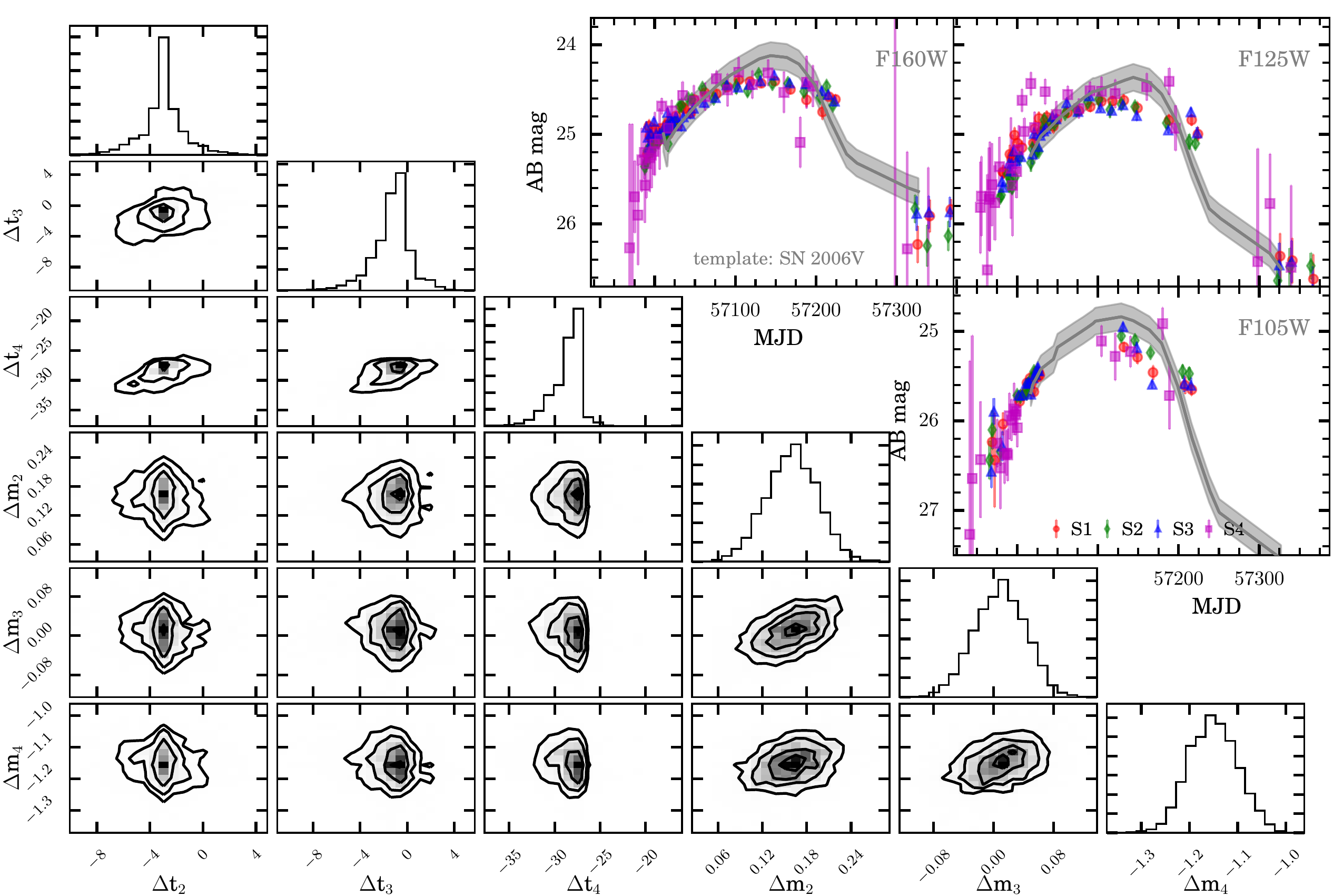}
   \caption{Results of the template fits to the SN Refsdal light
     curves using the best-matching template, which is the SN
     1987A-like Type II SN 2006V.  Three panels in the upper right
     show the composite light curve from images S1--S4, after applying
     the time and magnitude shifts that bring S2--S4 into the frame of
     the reference light curve S1 and maximize the likelihood function
     (Equation~\ref{eqn:Likelihood}) for the SN 2006V template.  The
     SN 2006V template light curve is overplotted \change{as a solid
       grey line, with the shaded band indicating the assumed 0.15 mag
       model uncertainty.}  Panels in the lower left show 2-D
     marginalized probability contours for each of the \change{six fit
       parameters that define relative time delays and
       magnifications.} Contours are shown at 0.5, 1.0, 1.5 and
     2.0$\sigma$. Histograms at the top of each column show the 1-D
     marginalized probability distribution for the parameter
     corresponding to that column.}
   \label{fig:refsdal_template_fits}
\end{figure*}

Figure~\ref{fig:refsdal_template_fits} shows the maximum likelihood
light curve model, which is based on the SN 2006V template.  This
template effectively matches the general character of the SN Refsdal
light curve, with a slow rise to maximum followed by a sharp drop.
There are, however, notable systematic deviations, such as the
sharpness of the peak and the steepness of the drop-off in the F160W
band.  Figure~\ref{fig:refsdal_template_fits} also plots the 1-D and
2-D probability distributions for \change{the six free parameters in
  the model that set the relative time delays and magnifications
  (i.e., these panels do not show four ``nuisance'' parameters that
  set the date of peak brightness and the SN color).}  The templates
SN 1987A and SN 2000cb provide a similar quality of fit to the data,
as reflected in their posterior probabilities in
Table~\ref{tab:TemplateFitResults}.  The time delay and magnification
measurements from other models are broadly consistent, although they
are substantially less effective at matching the observed photometry.

\input{TableTemplateFitResults}

The composite mean and uncertainty for each parameter--derived from
the BMA method (Equation~\ref{eqn:BMA})--are reported in the final row
of Table~\ref{tab:TemplateFitResults}.  This locates the time of peak
for image S1 in the F160W band -- arbitrarily selected as our
  reference light curve -- to be MJD$_{pk}$=57138 (26 April, 2015),
with an uncertainty of $\pm$10 days. This parameter in particular
should be taken with caution, as the best-fitting model shown in
Figure~\ref{fig:refsdal_template_fits} is clearly mis-representing the
behavior of SN Refsdal near peak brightness.

\subsection{Maximally Constrained Model Fit}\label{ssec:MaximallyConstrainedModelFit}

The fitting procedure described above is limited insofar as it only
employs the rest-frame $B, V$ and $R$ bands. This is a necessary
restriction, as the majority of known SN 1987A-like events do not have
extensive observations in rest-frame ultraviolet bands that could be
used to fit the F814W, F606W and F435W observations of SN Refsdal.
Furthermore, we have left out an important physical constraint, by
allowing the color of each template to be completely free, with a
separate parameter $C_\lambda$ shifting each bandpass independently.

 To derive a more physically constrained fit to the SN Refsdal light
 curves, we followed a prescription similar to that shown by
 \cite{Taddia:2012}.  We adopt the SN~1987A template for this purpose,
 as it has the most complete coverage in both wavelength and time, and
 in Section~\ref{ssec:TemplateFittingResults} we have seen that it is
 one of three models that can provide an adequate fit to the
 rest-frame $BVR$ light curve.  We first correct the SN 1987A template
 for host extinction using $E(B-V)$=0.16 mag \citep{Fitzpatrick:1990}
 and $R_V$=4.5 \citep{DeMarchi:2014} as appropriate for 30
   Doradus, the star-forming region within the Large Magellanic Cloud
 where SN~1987A exploded.  We then ``zero out'' the peak colors of the
 SN~1987A model by applying separate magnitude shifts in each band at
 the epoch of the rest-frame $R$ band peak brightness (i.e., forcing
 $B-V\equiv0$, $V-R\equiv0$, etc.). We apply the same shift across the
 light curve, so that the color evolution of the model still matches
 the observed color curve of SN 1987A. We then apply a
 temperature-based color-correction following
\begin{equation}
m_{\lambda}-V=-2.5\log\biggl[\int B_{\lambda}(T)\,S_{\lambda}\,\lambda\,d\lambda\biggr]+ C_{\lambda, AB}-C_{V, AB},
\end{equation}
\noindent where the magnitude correction for a given passband,
$m_{\lambda}$, is defined by the color-correction relative to the
$V$-band, which is the product of the filter throughput,
$S_{\lambda}$, and the Planck function at a given temperature,
$B_{\lambda}(T)$. $C_{\lambda, AB}$ and $C_{V, AB}$ define the system
response through the given passband and the $V$-band.

\begin{figure*}[ht] 
   \centering
   \includegraphics[width=\textwidth]{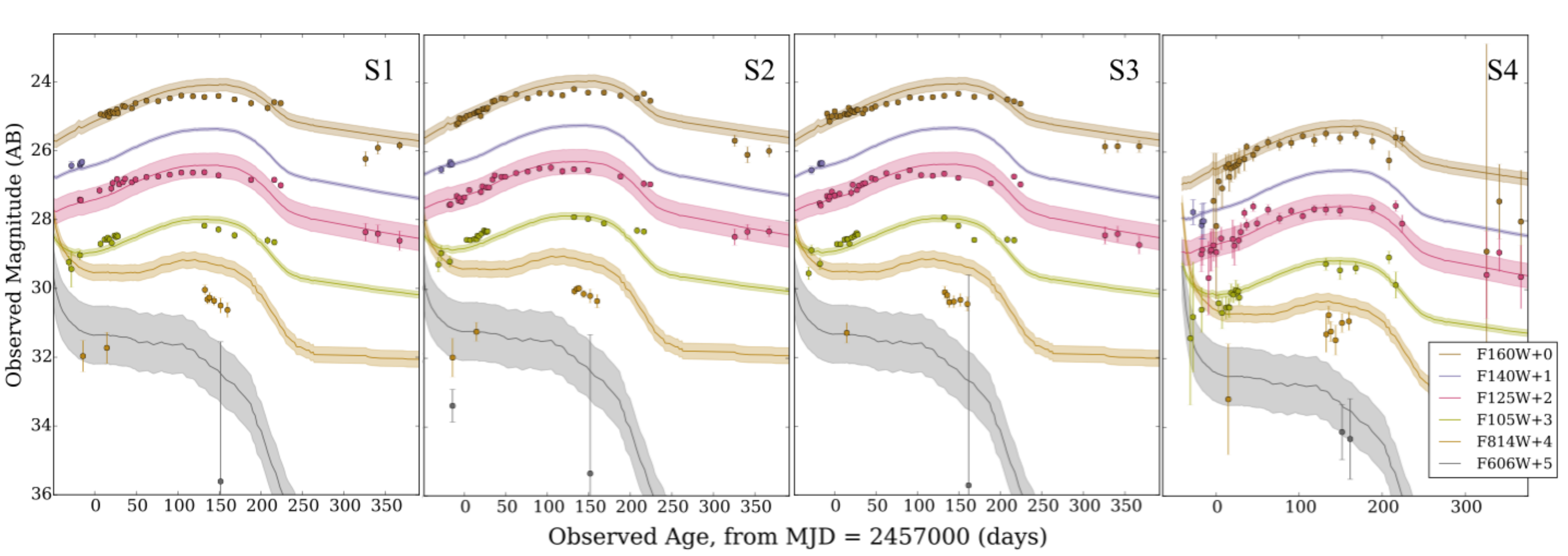}
   \caption{\footnotesize Observed light curves for the four
     SN~Refsdal images, S1--S4, as labeled, with vertical offsets as
     indicated in the legend. Curves show the ``maximally
     constrained'' template fit, based on the SN~1987A light curves,
     color corrected to match a peak blackbody temperature of $T=5300$
     K. Shaded bands indicate uncertainties from template photometric
     error and uncertainties in the k-correction.}
   \label{fig:MaximallyConstrainedFit}
\end{figure*}

Figure~\ref{fig:MaximallyConstrainedFit} shows the maximum likelihood
fit of this revised SN 1987A model to the observed SN Refsdal data in
all bands (including the optical bands from ACS/WFC).  From this fit
we find that the color--temperature of SN~Refsdal around maximum light
is $T\approx5300\,K$.  This is consistent with the range of
temperatures ($\sim$4000-9000 K) seen for other SNe~1987A-like events
at the same epoch~\citep{Pastorello:2005,Pastorello:2012,Taddia:2012}.
As discussed in \citet{Kelly:2015c}, the SN~Refsdal rest-frame optical
colors (with $B-V\approx0.5$ mag) are slightly bluer than most
SN~1987A-like objects. This blue color is most consistent with more
luminous SN~1987A analogs, such as SNe~2006V and
2006au~\citep{Taddia:2012}.

This modified SN~1987A model is not as good a fit to the data as the
best-fit SN 2006V model with completely free color terms from
Section~\ref{ssec:TemplateFittingResults}. This may be the result of
poorer matching of SN~1987A-like SEDs to smooth blackbody spectra in
bluer wavelengths, as was seen in \cite{Taddia:2012}, presumably due
to atmospheric line blanketing in these events. There is a solid
physical basis for the idea that color differences in the class of SN
1987A-like explosions stem primarily from differences in their
photospheric temperature.  These temperature differences may arise
from different explosion energies driven by a diversity in progenitor
masses. However, temperature differences alone can not explain the
wide diversity of this SN sub-class, indicating that other physical
parameters also strongly influence the color and color evolution.

Keeping these caveats in mind, we can nevertheless derive alternative
constraints on gravitational lensing parameters from this
color-temperature-corrected SN 1987A model.  We find again a set of broadly
consistent time delay and magnification estimates: $\Delta
t_{S2:S1}=-1.0\pm1.2$ days, $\Delta t_{S3:S1}=0.4\pm1.1$ days, and
$\Delta t_{S4:S1}=14.1\pm2.9$ days; $\mu_{\rm S2}/\mu_{\rm
  S1}=1.14\pm0.07$, $\mu_{\rm S3}/\mu_{\rm S1}=1.05\pm0.07$, and
$\mu_{\rm S4}/\mu_{\rm S1}=0.34\pm0.09$. The uncertainties here
reflect only statistical error estimates, inferred from the
photometric and model errors.

%% file: TableTemplateFitResults.tex

\begin{deluxetable*}{ll c ccc ccc}
  \tablecolumns{11}
  \tablecaption{\sc Time Delay and Magnification Ratio Measurements from SN Light Curve Template Fitting \label{tab:TemplateFitResults}}
  \tablehead{
    \colhead{Model} & \colhead{$p(T_k|{\bf D})$} & \colhead{MJD$_{\rm pk}$} & \colhead{$\Delta t_{\rm S2:S1}$} & \colhead{$\Delta t_{\rm S3:S1}$} & \colhead{$\Delta t_{\rm S4:S1}$} & \colhead{$\mu_{\rm S2}/\mu_{\rm S1}$} & \colhead{$\mu_{\rm S3}/\mu_{\rm S1}$} & \colhead{$\mu_{\rm S4}/\mu_{\rm S1}$} \\
  }
  \startdata    
SN 1987A   &  0.21   & 57148.0$^{+3.2}_{-1.8}$  & 7.58$^{+1.59}_{-3.52}$ & 5.45$^{+1.33}_{-4.13}$ & 20.80$^{+1.99}_{-4.51}$ & 1.127$^{+0.031}_{-0.031}$ & 1.019$^{+0.028}_{-0.028}$ & 0.331$^{+0.012}_{-0.012}$ \\[2mm] 
SN 1998A   &  1e-09  & 57169.0$^{+1.9}_{-2.1}$  & 6.99$^{+1.40}_{-4.16}$ & 5.87$^{+1.93}_{-4.98}$ & 28.87$^{+3.09}_{-2.09}$ & 1.159$^{+0.032}_{-0.032}$ & 1.038$^{+0.038}_{-0.038}$ & 0.380$^{+0.018}_{-0.018}$ \\[2mm] 
SN 2000cb  &  0.34   & 57125.0$^{+1.1}_{-2.0}$  & 1.98$^{+3.31}_{-1.33}$ & -0.41$^{+2.74}_{-0.98}$ & 19.69$^{+2.68}_{-1.17}$ & 1.138$^{+0.031}_{-0.031}$ & 1.000$^{+0.028}_{-0.028}$ & 0.344$^{+0.016}_{-0.016}$\\[2mm] 
SN 2006V   &  0.44   & 57143.8$^{+0.8}_{-1.8}$  & 3.01$^{+1.24}_{-1.23}$ & 1.03$^{+1.70}_{-0.94}$ & 28.09$^{+2.21}_{-0.70}$ & 1.159$^{+0.032}_{-0.043}$ & 1.009$^{+0.028}_{-0.037}$ & 0.347$^{+0.016}_{-0.016}$ \\[2mm] 
SN 2006au  &  3e-05  & 57161.2$^{+0.3}_{-0.1}$  & 6.24$^{+0.30}_{-0.28}$ & 3.33$^{+0.35}_{-3.06}$ & 31.19$^{+0.78}_{-0.87}$ & 1.127$^{+0.031}_{-0.031}$ & 1.000$^{+0.037}_{-0.028}$ & 0.384$^{+0.018}_{-0.018}$ \\[2mm] 
SN 2009E   &  3e-05  & 57160.2$^{+1.6}_{-1.5}$  & 4.35$^{+1.93}_{-2.06}$ & 2.97$^{+1.75}_{-2.35}$ & 21.85$^{+3.16}_{-1.98}$ & 1.107$^{+0.031}_{-0.031}$ & 1.000$^{+0.028}_{-0.028}$ & 0.353$^{+0.013}_{-0.016}$ \\[2mm] 
BMA Mean\tablenotemark{a}  & \nodata & 57138$\pm$10 & 4$\pm$4  & 2$\pm$4 & 24$\pm$5 & 1.15$\pm$0.05 & 1.01$\pm$0.04 & 0.34$\pm$0.02 \\ 

 \enddata
\tablenotetext{a}{Mean value for each parameter, computed using the Bayesian Model Averaging method (see text for details).}
\end{deluxetable*}

%% file: FlexFitting.tex
\section{Time Delay Measurements with Flexible Light Curve Models}
\label{sec:FlexFitting}

As Figure~\ref{fig:refsdal_template_fits} shows, even the best-fit
template-based model shows systematic residuals and does not provide a
good representation of the observed data. SN Refsdal is not quite a
clone of other observed 87A-like Type II SNe. Thus as a second approach for
measuring the time delays between the four Refsdal sources, we used a
series of flexible light curve models (splines and polynomials) to
represent the underlying light curve shape.  By adopting these
free-form curves in place of the rigid SN light curve templates, we
can derive time delays that are agnostic about the classification of
SN Refsdal.  This allows for the possibility that SN Refsdal is unlike
any of the available SN templates, and we may thereby avoid a
systematic bias that could be introduced by assuming an incorrect
light curve shape. The cost of this more flexible approach is that we
lose the physical/empirical priors on the light curve shape and color
that a well-matched template would afford.  This second
  approach is therefore much closer to the methodology typically used
  for measuring lensed quasar time delays
  \citep[e.g.][]{Tewes:2013a,Liao:2015}, where there is no way to
  apply an informative prior for the intrinsic light curve shape.

The first year of the SN~Refsdal light curve is fundamentally very
simple: a slow rise to a broad peak, and a gradual decline.  
To approximate this intrinsic light curve shape with the
  simplest possible functional form, we start by adopting a low-order
Chebyshev polynomial of the first kind, which gives the magnitude of
image $i$ in band $j$ at time $t$ as:
\begin{multline}
  m_{i,j}(t) = c_0 + c_1 T_1(t + \Delta t_i) + \\ 
      c_2 T_2(t + \Delta t_i) + [...] + \Delta m_i 
\label{eqn:Chebyshev}
\end{multline}  
\noindent where the coefficients $c_n$ are free parameters in the
model and the polynomial components $T_n$ are defined by the Chebyshev
recurrence relation $T_{n+1}(x) = 2xT_n(x) - T_{n-1}(x)$,
with $T_0(x)=1$ and $T_1(x)=x$. The effects of gravitational lensing
are represented in separate time shifts $\Delta t_i$ and magnitude
shifts $\Delta m_i$ for each of the four images, though we fix image
S1 as our reference point by setting $\Delta t_1\equiv0$ and $\Delta
m_1\equiv0$ (i.e., we are only fitting for {\it relative} time delays
and magnifications).  Note that the intrinsic color of the SN
  is accounted for by having a separate polynomial fit to each band.
When fitting this model to the SN Refsdal data, we use only the
F160W, F125W and F105W bands, for which we have sufficient
  data to effectively constrain the peak of the light curve
  independently in each band.  All together, this means a 2$^{\rm
  nd}$-order polynomial model has 15 free parameters: three
polynomial coefficients in each of three bands to define the light curve
shape and color, and six parameters for the time delays and
  magnifications of S2-S4 relative to S1. Increasing the degree of
the polynomial by one adds three additional free parameters (one new
polynomial coefficient for each passband).

To allow for more complex intrinsic light curve
  shapes, we also evaluated cubic spline fits, using one, two and three
internal spline knots at fixed positions along the time axis. The
knots were arbitrarily set to MJD = [57150] for the single-knot
spline, [57000, 57200] for the two-knot spline, and [57000, 57100,
  57200] for the three-knot spline.

\begin{figure*}[p!]
   \centering
   \includegraphics[width=\textwidth]{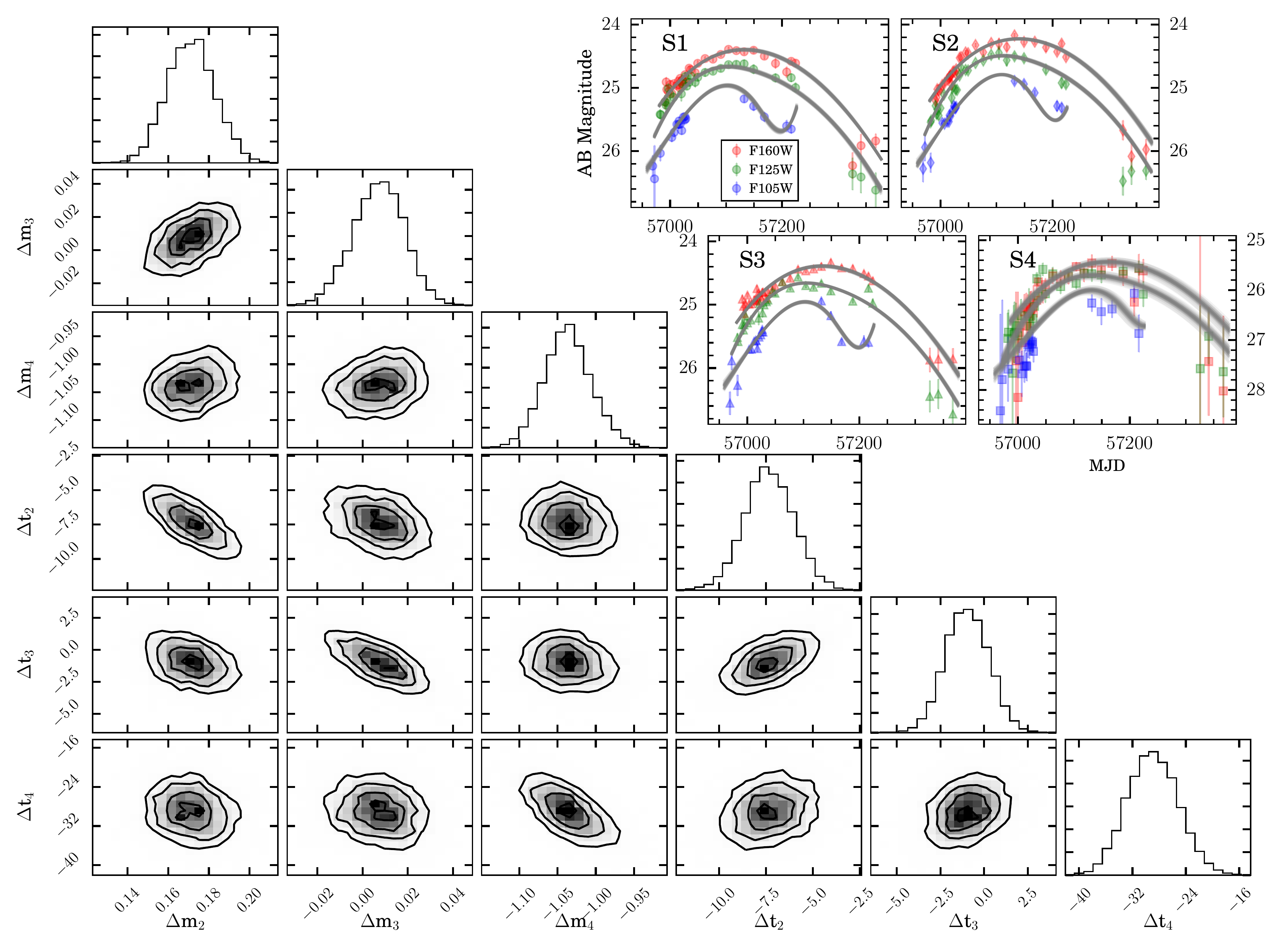}
   \caption{ Results of fitting the SN Refsdal light curves using a
     cubic spline with a single internal knot. Four panels in the
     upper right show the spline fits to the observed data, with each
     panel showing a single Refsdal image (S1--S4, as labeled). The
     F160W, F125W and F105W data are plotted together as red, green
     and blue points, respectively. Overlaid grey curves show the
     optimized spline functions, which are fit simultaneously to all
     four images. \change{Each grey band comprises a sample of 100
       curves drawn randomly from the MCMC chain, to give an
       indication of the range of variation in the shapes of curves
       that have parameters close to their optimal values.}  Panels in
     the lower left show marginalized 2-D posterior probability
     distributions for each of the 6 lensing parameters (magnitude
     shifts and time delays relative to the reference light curve,
     S1).  As in Figure~\ref{fig:refsdal_template_fits}, contours mark
     the 0.5, 1.0, 1.5, and 2.0$\sigma$ confidence regions, and
     histograms at the top of each column show 1-D marginalized
     posterior probability distributions.
     \label{fig:SplineFit1}}
\end{figure*}

To check whether this arbitrary knot placement could bias the
measurement of lensing parameters, we also evaluated the use of more
sophisticated algorithms for optimizing the number and location of
internal cubic spline knots.  We used tools from the \pycs
\citep{Tewes:2013a}\footnote{\url{http://obswww.unige.ch/~tewes/cosmograil/public/pycs/index.html}}
and \SNooPy software packages
\citep{Burns:2011,Burns:2015}.\footnote{\url{http://csp.obs.carnegiescience.edu/data/snpy}}
The \pycs program was orginally developed by the Cosmological
Monitoring of Gravitational Lenses (COSMOGRAIL) collaboration
\citep{Eigenbrod:2005}\footnote{\url{http://cosmograil.epfl.ch}} for
the measurement of gravitational lensing time delays from
single-filter quasar light curve sets. To collapse the multi-band SN
Refsdal light curves into a form suitable for use with \pycs, we used
a combination of the F125W and F160W observations, which were
collected concurrently in almost every epoch, and together have the
most complete and well-sampled coverage of the Refsdal light curve.
The \SNooPy software suite was developed by the Carnegie Supernova
Project \citep{Hamuy:2006} to provide general purpose SN light curve
fitting tools, especially for Type Ia SNe.  The \SNooPy spline fitting
tools automatically enforce a restriction on the flexibility of the
spline curve model by using the ``hyperspline'' algorithm
\citep{Thijsse:1998}.  This method is designed to find a spline
representation for noisy data without allowing the spline to follow
every noise feature. This is achieved by starting with an
interpolating spline (one knot at every observed data point) and
iteratively removing knots to optimize the Durbin-Watson statistic
\citep{Durbin:1950,Durbin:1951}, which tests for serial correlation in
the least squares regression. When applying \snpy we defined a
separate spline curve for the F160W, F125W and F105W bands.  Once
again we found that the added flexibility and optimal placement of
spline knots in these two packages did not lead to any significant
changes in the inferred time delays or magnifications.

\subsection{Flexible Curve Fitting Results}\label{ssec:FlexFitResults}

\input{TableFlexFitResults}

Table~\ref{tab:FlexFitResults} reports the time delays and magnitude
shifts of the sources S2--S4 relative to S1, derived from the
polynomial and spline fits described above. The best model,
  as measured by the total posterior probability $p(M_k|{\bf D})$, is
  a cubic spline with a single internal knot.
  Figure~\ref{fig:SplineFit1} shows this best-fit spline model, along
  with marginalized posterior probability distributions for each of
  the lensing parameters. All other models except
  the $2^{\rm nd}$-order Chebyshev polynomial provide a similar
  quality of fit to the data.  Furthermore, the relative time delays
  and magnifications inferred from all models are quite consistent.
  As in Section~\ref{sec:LightCurveTemplateFitting}, we use the BMA
  method to combine the parameter estimates from all of these models,
  deriving the values given in the final row.  These measurements are
  fully consistent within the uncertainties with the values inferred
  from SN light curve template fitting.

To explore whether the light curve can be effectively described with
fewer parameters, we also evaluated a set of ``minimalist'' polynomial
and spline models. In this case we assume that all bands (F160W, F125W
and F105W) have the same intrinsic light curve shape, meaning that
they would reach peak brightness at the same epoch.  This is not a
good assumption for SN 1987A-like explosions, which tend to reach peak
brightness much earlier in bluer bands
\citep[e.g.][]{Pastorello:2012,Taddia:2012}.  Using both the net
posterior probability and the Bayesian Information Criterion
\citep[BIC,][]{Schwarz:1978} as metrics to evaluate the fitness of
these simpler models, we found that we consistently get a better fit
when using the more complex alternative. That is, using three separate
polynomials or splines to describe each band independently gives a
better representation of the light curve shape, regardless of the
degree of the polynomial or the number of knots in the spline.
Expanding the input data to include the F140W and F814W observations
does not change these conclusions.

\subsection{Uncertainty Estimates from Mock Light Curves}\label{ssec:FlexFitUncertaintyEstimates}

The uncertainty estimates given in Table~\ref{tab:TemplateFitResults}
and \ref{tab:FlexFitResults} reflect the statistical uncertainties due
to photometric measurement error.  The error in the BMA mean also
accounts for some of the systematic errors that may be introduced by
adopting an inappropriate functional form to describe the intrinsic
light curve shape.  As an alternative means to estimate systematic
uncertainties, we follow the algorithm of \citet{Tewes:2013a},
generating 1,000 mock light curves from the best-fit
  polynomial and spline curve fits, after introducing artificial time
delays and magnifications drawn from uniform distributions about the
best-fit values. We then fit the mock curves with the same procedures
described above and measure the difference between the input and
recovered values of the time delays and magnifications.

Each mock light curve is constructed with observations at the actual
dates and in the same filters where SN Refsdal was observed.  For each
mock data point we start with the exact magnitude predicted by the
best-fit model (polynomial or spline) for that epoch.  Then we add a
magnitude offset $\Delta m_{\rm noise}$ drawn from a normal
distribution.  If the observed data point was within 1.5$\sigma$ of
the model, then we set the standard deviation of that normal
distribution equal to the photometric error of that observed data
point.  For data points where the difference between the observation
and the best-fit model is $>1.5\sigma$, we set the standard deviation
equal to the residual between the observation and the model, and we
preserve the sign of the residual when applying the offset.  This
ensures that our mock light curves have a similar mix of random
offsets due to photometric error and correlated systematic offsets due
to possible mismatches between the true light curve shape and the
best-fit model.  However, the runs of data points with similar
residuals appear at different phases relative to the light curve peak
for each mock light curve, because we have introduced random time
delay shifts for the mock S1-S4 events.

\begin{figure*}[htb]
   \centering
   \includegraphics[width=\textwidth]{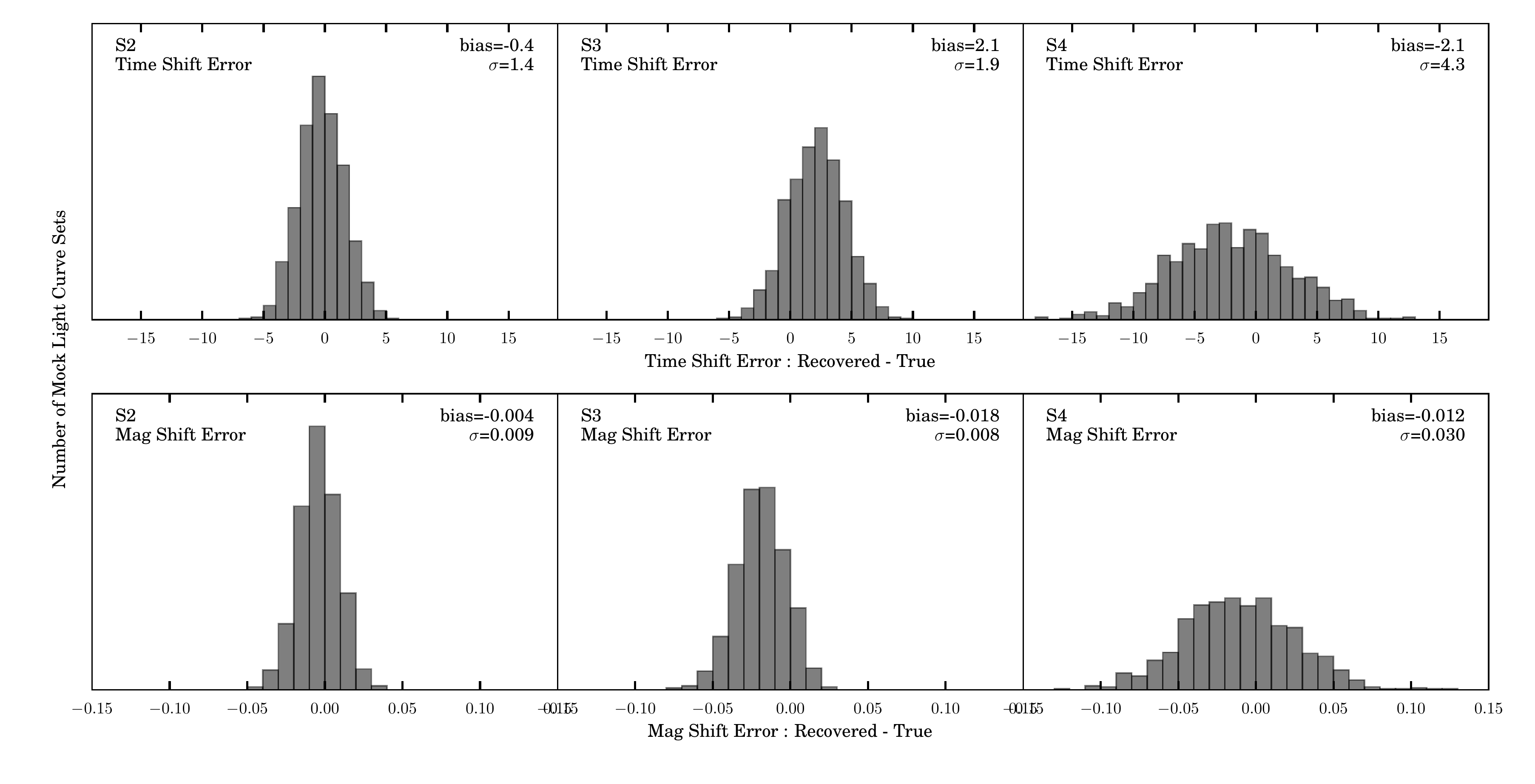}
   \caption{ Histograms from the mock light curve analysis,
     using a single-knot cubic spline to represent the
     intrinsic light curve shape.  Three panels in the top row
     illustrate the distributions of time delay errors (in days) as
     the recovered time delay minus the input (true) time delay for
     each mock light curve. The bottom row shows the same difference,
     but for the magnitude shift (in mags) accounting for the lensing
     magnification of each mock light curve.
   \label{fig:ChebyshevHistogram}}
\end{figure*}

Figure~\ref{fig:ChebyshevHistogram} shows histograms of the
``recovered-minus-actual'' time delays and magnitude shifts from this mock
light curve analysis.  Fitting a Gaussian to each histogram, we report
the mean and standard deviation in each panel.  We then adopt the
standard deviation of the Gaussian fit as the statistical uncertainty
for each lensing parameter.  A non-zero value in the mean reflects a
potential systematic bias in the fitting procedure, and we include
this in our total uncertainty estimate, adding in quadrature to the
statistical uncertainty.

%% file: TableFlexFitResults.tex
\begin{deluxetable*}{lcc ccc ccc}
  \tablecolumns{9}
  \tablecaption{\sc Time Delay and Magnification Ratio Measurements from Polynomial and Spline Fits \label{tab:FlexFitResults}}
  \tablehead{
  \colhead{Model} & \colhead{$p(M_k|{\bf D})$} & \colhead{MJD$_{\rm pk}$}\tablenotemark{a} & \colhead{$\Delta t_{\rm S2:S1}$} & \colhead{$\Delta t_{\rm S3:S1}$} & \colhead{$\Delta t_{\rm S4:S1}$} & \colhead{$\mu_{\rm S2}/\mu_{\rm S1}$} & \colhead{$\mu_{\rm S3}/\mu_{\rm S1}$} & \colhead{$\mu_{\rm S4}/\mu_{\rm S1}$} \\
  }
  \startdata    
Chebyshev, deg=2 & 0.02  & 57136.7     & 8.3 $\pm$1.5 & -2.1 $\pm$1.6 & 32.5 $\pm$4.4 & 1.17 $\pm$0.01 & 0.99 $\pm$0.01 & 0.39 $\pm$0.01 \\
Chebyshev, deg=3 & 0.13  & 57133.8     & 6.1 $\pm$1.5 &  0.2 $\pm$1.4 & 30.6 $\pm$3.6 & 1.17 $\pm$0.01 & 1.00 $\pm$0.01 & 0.39 $\pm$0.01 \\
Chebyshev, deg=4 & 0.19  & 57133.5     & 6.7 $\pm$1.4 &  1.0 $\pm$1.3 & 24.7 $\pm$3.4 & 1.17 $\pm$0.01 & 1.00 $\pm$0.01 & 0.38 $\pm$0.01 \\
Chebyshev, deg=5 & 0.19  & 57133.4     & 6.5 $\pm$1.4 &  1.0 $\pm$1.3 & 24.2 $\pm$3.4 & 1.17 $\pm$0.01 & 1.00 $\pm$0.01 & 0.38 $\pm$0.01 \\
Spline, 1 knots  & 0.19  & 57132.7     & 7.4 $\pm$1.4 &  1.0 $\pm$1.3 & 29.0 $\pm$3.9 & 1.17 $\pm$0.01 & 1.01 $\pm$0.01 & 0.38 $\pm$0.01 \\
Spline, 2 knots  & 0.15  & 57131.5     & 8.5 $\pm$1.3 &  0.4 $\pm$1.4 & 31.3 $\pm$3.9 & 1.18 $\pm$0.01 & 1.00 $\pm$0.01 & 0.39 $\pm$0.01 \\
Spline, 3 knots  & 0.13  & 57127.0     & 7.4 $\pm$1.2 & -0.1 $\pm$1.2 & 23.6 $\pm$3.4 & 1.17 $\pm$0.01 & 1.00 $\pm$0.01 & 0.38 $\pm$0.01 \\
BMA mean       & \nodata & 57132$\pm$3 & 7$\pm$2      &  0.6 $\pm$2   & 27$\pm$6      & 1.17 $\pm$0.02 & 1.00 $\pm$0.01 & 0.38 $\pm$0.02 \\
\enddata \tablenotetext{a}{The date of peak brightness inferred for
  the reference curve S1 in the F160W band. Note that this is not a
  singular parameter in these light curve models, but rather is
  accounted for in the coefficients of the polynomial or spline curve
  functions. We report here the value derived from locating the peak
  of the maximum likelihood model.}
\end{deluxetable*}

%% file: Discussion.tex
\section{Summary and Discussion}\label{sec:Discussion}

The light curve of SN Refsdal after one full year of \HST imaging
observations reveals this to be an unusual object.  The extremely
broad shape of the light curve in rest-frame optical bands is
incompatible with the relatively rapid rise of normal Type I and Type
II SNe. The observed shape has distinct similarities with SN 1987A,
and in a companion paper we have concluded that SN Refsdal
is most likely a member of the peculiar Type II sub-class defined by
the SN 1987A proto-type \citep{Kelly:2015c}.

\input{TableResultsSummary}

In this work we have explored two methods for measuring time delays
from the SN Refsdal light curves. We first used a set of six SN light
curve templates with shapes similar to SN 1987A, allowing the template
colors to float as free parameters to accommodate the very blue
rest-frame optical colors of SN Refsdal.  We then adopted flexible
polynomial functions as an alternative description of SN Refsdal's
intrinsic light curve shape.  We find that the SN can be well
represented by a very simple set of low-order Chebyshev polynomials or
cubic splines, and from these fits we derive consistent results for
the relative time delays and magnifications.  The results from these
complementary time delay measurement strategies are summarized in
Table~\ref{tab:ResultsSummary}.

Each method independently provides measurements of the time delays for
S2, S3 and S4 (relative to S1) with a precision of $\pm$2 to 8
days. This level of precision is promising, as it suggests that a
similarly cadenced monitoring campaign could deliver a relative
precision of $\sim$1\% on the time delay to the next image, SX,
expected to reach peak brightness approximately one year after the
observed S1 peak \citep{Kelly:2015d}.  Similarly, the magnification
ratios relative to S1 are measured to better than 3\% precision for S2
and S3, and better than 10\% for S4.

\subsection{Comparison to Model Predictions}

As the first multiply-imaged SN ever seen, SN Refsdal has garnered
great interest from the lens modeling community.  The
  presentation of the SN Refsdal discovery by \citet{Kelly:2015a}
  included a first analysis of the SN lensing, using the ``light
  traces mass'' (LTM) modeling approach \citep{Broadhurst:2005,Zitrin:2009a}.
  Within a week of the initial announcement, two other teams produced
  revised strong lensing models for the \macs1149 cluster
  \citep{Oguri:2015,Sharon:2015}.  These were tuned to give more
accurate predictions for the magnifications and time delays of the
Einstein Cross images, as well as for the SX image.
Subsequently, other lens modeling groups have produced updated models,
taking advantage of improved imaging and spectroscopic data on this
field to generate models using better catalogs of multiply-imaged
galaxies, as well as better precision in reproducing those
strong-lensing constraints \citep{Diego:2015b,Jauzac:2015c}.  Most
recently, \citet{Treu:2015b} presented a coordinated effort from five
lens modeling groups to produce new \macs1149 lens models with a set
of collectively vetted strong lensing constraints. The individual
results from each group are being published separately, providing
details on each group's modeling approach
(\citealt{Grillo:2015b}; \citealt{Kawamata:2015}; Diego et al., in preparation; Sharon et
  al., in preparation; Zitrin et al., in preparation).  A primary goal of this
effort was to examine how different choices and assumptions in the
modeling methodology can affect the predictions of magnifications and
time delays for a strongly-lensed source such as Refsdal.

\begin{figure}
    \centering
    \includegraphics[width=\columnwidth]{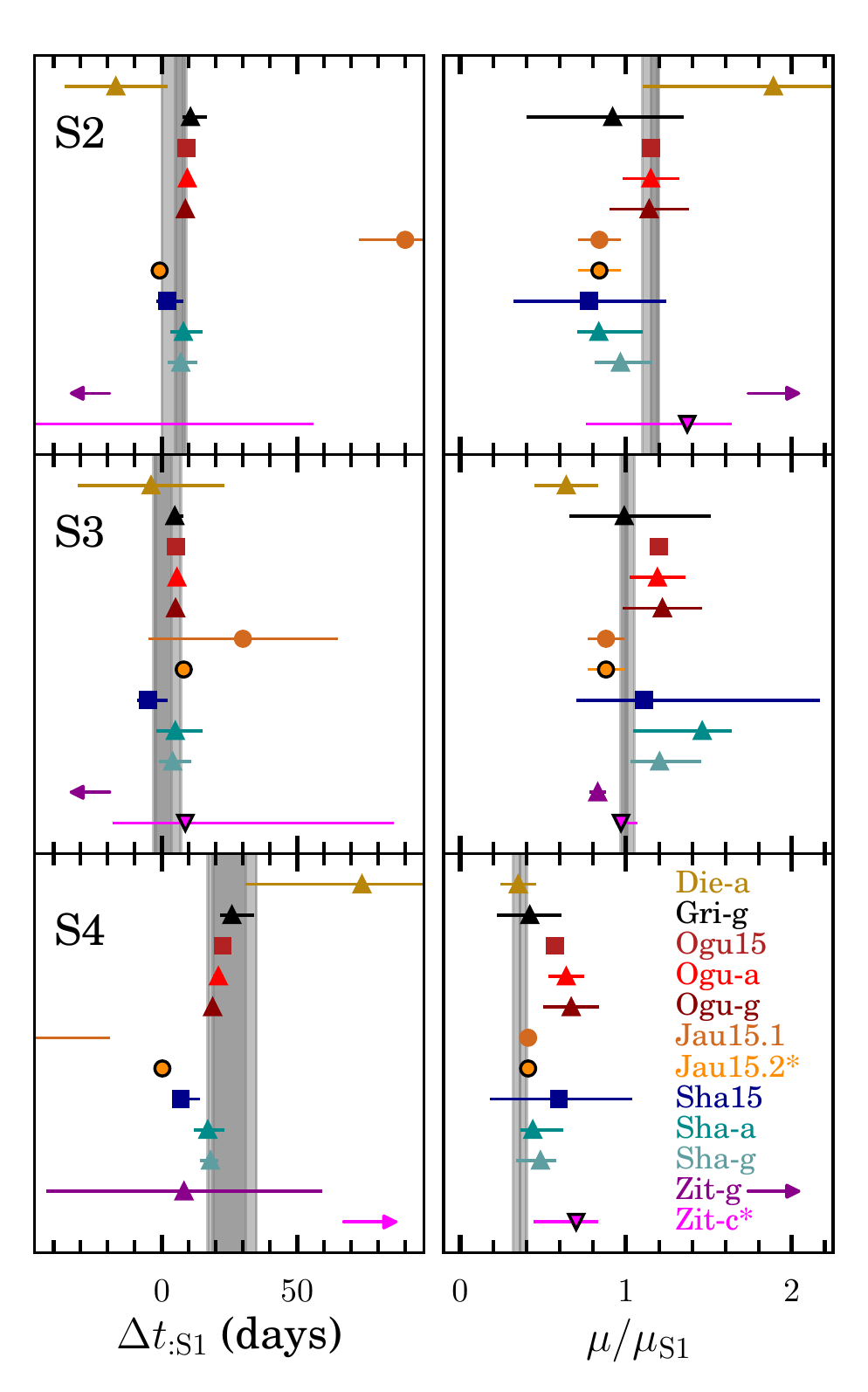}
    \caption{Comparison of the measured time delays against values
      predicted by lensing models. The three rows of panels present
      results for images S2, S3, and S4 from top to bottom.  Panels in
      the left column plot time delays and in the right column they
      show magnification ratios (relative to S1 in both cases).
      Vertical gray bars indicate the measurements from
      Table~\ref{tab:ResultsSummary}. The darker shaded regions
      indicate overlapping measurements from the two methods presented
      in Sections~\ref{sec:LightCurveTemplateFitting} and
      \ref{sec:FlexFitting}.  Predicted time delays from published
      lens models are plotted as colored points, using the key given
      in the lower right panel (see text for details).  \change{The
        two ``unblind'' models are marked with asterisks and
        plotted with black outlines.}  Arrows indicate points that fall
        outside the plotted range.
    \label{fig:MeasurementsVsModels}}
 \end{figure}

Figure~\ref{fig:MeasurementsVsModels} presents a comparison of our
measured magnification ratios and time delays for images S1$-$S4
against all published lens model predictions available as of December
15, 2015.  The earliest models, first posted within a week of
  the SN Refsdal discovery, are shown as squares
  \citep{Oguri:2015,Sharon:2015}.\footnote{Note that
    \citet{Oguri:2015} did not report uncertainties for the Ogu15
    model.}  The recently updated model of \citet{Jauzac:2015c} is
  plotted with circles, and the set of models from the
  \citet{Treu:2015b} model comparison program (Die-a, Gri-g, Ogu-a, Ogu-g, Sha-a, Sha-g, Zit-g, Zit-c) are plotted with
  trangles.

\change{For all but two of these models,} the time delay and
magnification ratio measurements presented here were not available for
use as input strong-lensing constraints, or as an intermediate check
to guide the model development. \change{Thus, the comparison of our
  measurements against these models is effectively a true blind test
  of the predictive power of each model.  The two exceptions are the
  Zit-c model and the Jau15.2 model, which were updated after the
  initial release of these SN measurements. These ``unblind''
  models are marked by asterisks and plotted with a black outline in
  Figure~\ref{fig:MeasurementsVsModels}.}

  \change{The Zit-c model} is a corrected version of the Zit-g model,
  updated after the SN measurements presented in this work were
  known.  The Zit-c model does not use any measured time delays or
  magnification ratios as input constraints, but does use the
  previously known positions of the S1-S4 images (as do all the models
  evaluated here).  The key change in the Zit-c model is that it
  allows the total mass of the lens galaxy to be a free parameter,
  which ensures that its critical curves pass through the Einstein
  cross (S1-S4), as required by those positional constraints.  For
  computational efficiency, both versions of this model were computed
  using a relatively low resolution grid \change{(0.065"/pix),} and
  the predictions of the model with respect to the S1--S4 time delays
  are therefore limited by this grid scale.  \change{The main source of
    difference between the predictions of the Zit-g and
    Zit-c models (which both use the LTM approach) and the
    predictions by other (analytic) models is the different
    parametrization of the mass distribution.}  See Treu et
  al. (2015a) for further details.

\change{The Jau15.1 model is the version presented by
  \citet{Jauzac:2015c} (first appearing in arXiv eprint v3), which
  adopts the model-predicted positions for the SN sources S1--S4 when
  computing the time delays.  These model-predicted source locations
  do not match the observed locations, leading to biases in the
  predicted time delays.  After the initial release of the SN
  measurements, a revision of \citet{Jauzac:2015c} (arXiv eprint v4)
  introduced the model labeled here as Jau15.2. This version instead
  uses the {\it observed} locations of images S1--S4 and adopts an
  analytic approach to compute the time delays that effectively forces
  S1--S4 to be spatially and temporally coincident at the source
  plane. These unblind time delays are presented as $\Delta t_{\rm
    CATS-src}$ by \citet{Jauzac:2015c}.}
  
 Although there are a few predictions that are discrepant \change{at
   $>2\sigma$}, the overall agreement between model predictions and
 our measurements is quite good. Unlike the situation for SN HFF14Tom,
 discussed in \citet{Rodney:2015a}, there is no indication of a
 consistent systematic bias among all models for any of the time
 delays or magnifications.  In the case of SN Refsdal, the
   Einstein Cross configuration is dominated by a single (galaxy)
   lens, which is a very different lensing regime than for a
   SN like HFF14Tom on the outskirts of the strong-lensing region of a
   cluster-scale lens. 

 \begin{figure}
    \centering
    \includegraphics[width=\columnwidth]{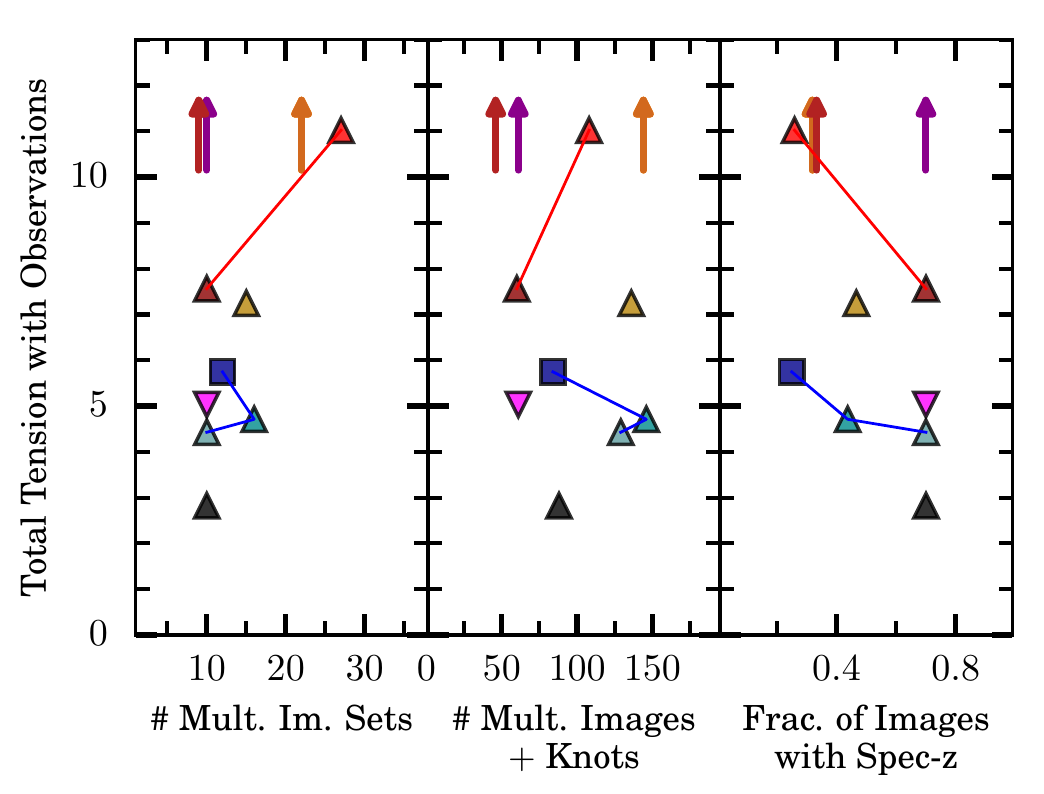}
    \caption{The tension between models and measurements,
        plotted against metrics assessing the quantity and quality of
        input strong lensing constraints.  The ordinate in all three
        panels marks the number of standard deviations separating the
        model predictions from the observations of time delays and
        magnification ratios, collapsed to a single value using
        Equation~\ref{eqn:Tension}. The abscissa in the left panel is
        the number of multiply imaged systems used as input
        constraints for each model, and in the middle panel it is the
        total number of images (counting all instances of each lensed
        galaxy), and includes all the distinct knots
        from the Refsdal host galaxy that are used as input positional
        constraints.  The right panel plots the tension against the
        fraction of multiply-imaged systems that have a spectroscopic
        redshift (a crude metric for ``quality'' of input constraints,
        following \citet{Rodney:2015b}).  Symbols and colors are as in
        Figure~\ref{fig:MeasurementsVsModels}. 
    \label{fig:Tension}}
 \end{figure}

The overall agreement between model predictions and observed time
delays is an encouraging indication of the accuracy of this current
generation of well-vetted models.  Furthermore, all of the most
up-to-date models
\citep{Grillo:2015b,Jauzac:2015c,Kawamata:2015,Treu:2015b} agree that
the date of peak brightness for the reappearance of SN
  Refsdal in image SX should occur within $1-1.5$
  years. That prediction is also broadly in agreement with earlier
models based on a more limited set of input data
\citep{Diego:2015,Kelly:2015a,Oguri:2015,Sharon:2015}.  A
  transient source at the expected position of image SX has now been
  detected, with magnitudes that are consistent with the
  magnification ratios and time delays predicted by several of these
  models \citep{Kelly:2015d}. As the full light curve of this new
  image is measured over the coming year, we will soon be able to complete
  this direct test of those falsifiable model predictions.

The natural experiment afforded by SN Refsdal gives us an
  opportunity to examine whether a particular modeling strategy or set
  of input constraints can deliver better estimates of the time delays
  and magnifications.  The SN observations should be particularly
  useful for identifying subtle systematic biases common to many
  models, as in \citet{Rodney:2015a}.
  To simplify this assessment, we define the ``total tension'' between a
  given model and the SN Refsdal measurements as

\begin{multline}
  \tau = \sum_{i=2-4} \frac{(\Delta t_i({\rm obs}) - \Delta t_i({\rm mod}))^2}{(\sigma^2_{t_i}({\rm obs}) + \sigma^2_{t_i}({\rm mod}))} +\\
  \frac{(\frac{\mu_i}{\mu_1}({\rm obs}) - \frac{\mu_i}{\mu_1}({\rm mod}))^2}{(\sigma^2_{\mu_i}({\rm obs}) + \sigma^2_{\mu_i}({\rm mod}))}
  \label{eqn:Tension}
\end{multline}
  
\noindent Figure~\ref{fig:Tension} plots this total tension
  against three metrics that quantify the input strong-lensing
  constraints used by each model: the number of multiple-image
  systems, the total number of images (including knots within the SN
  Refsdal host), and the fraction of multiply-imaged galaxies that
  have a spectroscopic redshift.  \citet{Rodney:2015a}
    performed a similar comparison, using the absolute magnification
    measurement from a lensed Type Ia SN to test the accuracy of 17
    lens models for the cluster Abell 2744.  That analysis suggested
    that simply increasing the {\it quantity} of strong lensing
    constraints did not in and of itself lead to a more
    accurate magnification prediction.  This SN Refsdal test
    reinforces that suggestion, as shown in the left and middle panels
    of Figure~\ref{fig:Tension}: the models using the greatest number
    of multiply-imaged systems tend to have a greater total tension
    with observations, and the same is true for those models with the
    greatest total number of images and knots from the SN Refsdal host
    galaxy.

\citet{Rodney:2015a} found evidence that the {\it quality} of
  input lensing constraints is mildly correlated with successful model
  predictions.  The rightmost panel of Figure~\ref{fig:Tension} again
  provides some support for this idea: the models that are most
  accurate in predicting the SN Refsdal time delays and magnifications
  all have a large fraction of strong-lensing constraints with
  spectroscopic redshifts (the same crude ``quality'' metric that was
  used by \citet{Rodney:2015a}).  Two of the model families from the
  \citet{Treu:2015b} comparison are particularly informative in this
  analysis.  The Sharon et al. model series (Sha15, Sha-a, Sha-g) and
  the Oguri et al. model series (Ogu15, Ogu-a, Ogu-g) were each
  generated by the same modeling team, using the same modeling
  toolkit, with the same basic model assumptions.  In
  Figure~\ref{fig:Tension} these sequences are plotted with connecting
  lines, and both follow the trend of increasing accuracy (lower
  tension) as the spec-$z$ fraction increases.  As such, these
  sequences provide an especially clean indication that the quality of
  strong-lensing constraints is a key ingredient for model accuracy.

When a lensed SN is not available for empirical tests of lens models,
it would be tempting to determine the best possible magnification or
time delay predictions from a set of independent models using a
``wisdom of the crowd'' approach.  For example, one might use a median
or an average from a set of models that includes one contribution from
every modeling team.  In the case of SN Refsdal, using such a method
to combine model predictions would in fact deliver an accurate and
precise estimate of the lensing time delays and magnifications.
However, this is not ideal. As discussed by, e.g.,
\citet{Treu:2015b}, the dispersion and bias of the predictions could
be overestimated by incorrect assumptions in some of the models, or
they could be underestimated if all models suffer from the same
incorrect assumption.  A more fruitful approach is to actually try to
understand why some models perform better than others, and what
assumptions are justified and what are not.

Additionally, one should note that the trends displayed in
  Figure~\ref{fig:Tension} are fairly weak, and the number of models
  being tested is quite small.  Thus, although these comparisons can
certainly be informative, we should be very cautious about making
inferences regarding the global fitness of these lens models based on
their precision or accuracy in predicting the measurable properties of
a single SN.  The value and limitations of such comparisons are
discussed in more detail by \citet{Rodney:2015a} and
\citet{Treu:2015b}, and the interested reader is referred there.

\subsection{Microlensing}\label{ssec:Microlensing}

Throughout this work, we have ignored the possible effects of {\it
  microlensing}: small-scale gravitational lensing perturbations due
to massive objects along the light path of any one image in the
quartet.  Instead, we have assumed that each of the S1--S4 light
curves is only affected by a single magnification factor that is
static over the duration of the light curve. It is, however, quite
possible that SN Refsdal is affected by either of two types of
microlensing.  The first is the traditional form of microlensing that
has been observed in lensed quasars \citep[e.g.][]{Kochanek:2004}.  In
this case, the effective transverse motion of stars in the lensing
galaxy changes the intervening lensing potential and causes
fluctuations in the light curve on a timescale of months or years
\citep[e.g.][]{Wyithe:2001,Schechter:2002,Schechter:2004}.
\citet{Dobler:2006} describe the second form of microlensing that is
unique to lensed SNe.  The SN light passing through the lensing plane
is distorted by a web of lensing potentials formed by all the
intervening stars in the lensing galaxy.  As the photosphere of the SN
expands, it intersects a larger section of this complex web, which can
result in microlensing fluctuations that affect the light curve on
timescales of weeks to months.  Analysis of such microlensing features
in a lensed SN light curve could potentially be used to make
inferences about the mass fraction and projected spatial density of
the stellar population in the lensing galaxy
\citep{Kolatt:1998,Dobler:2006}.

\begin{figure*}
   \centering
   \includegraphics[width=\textwidth]{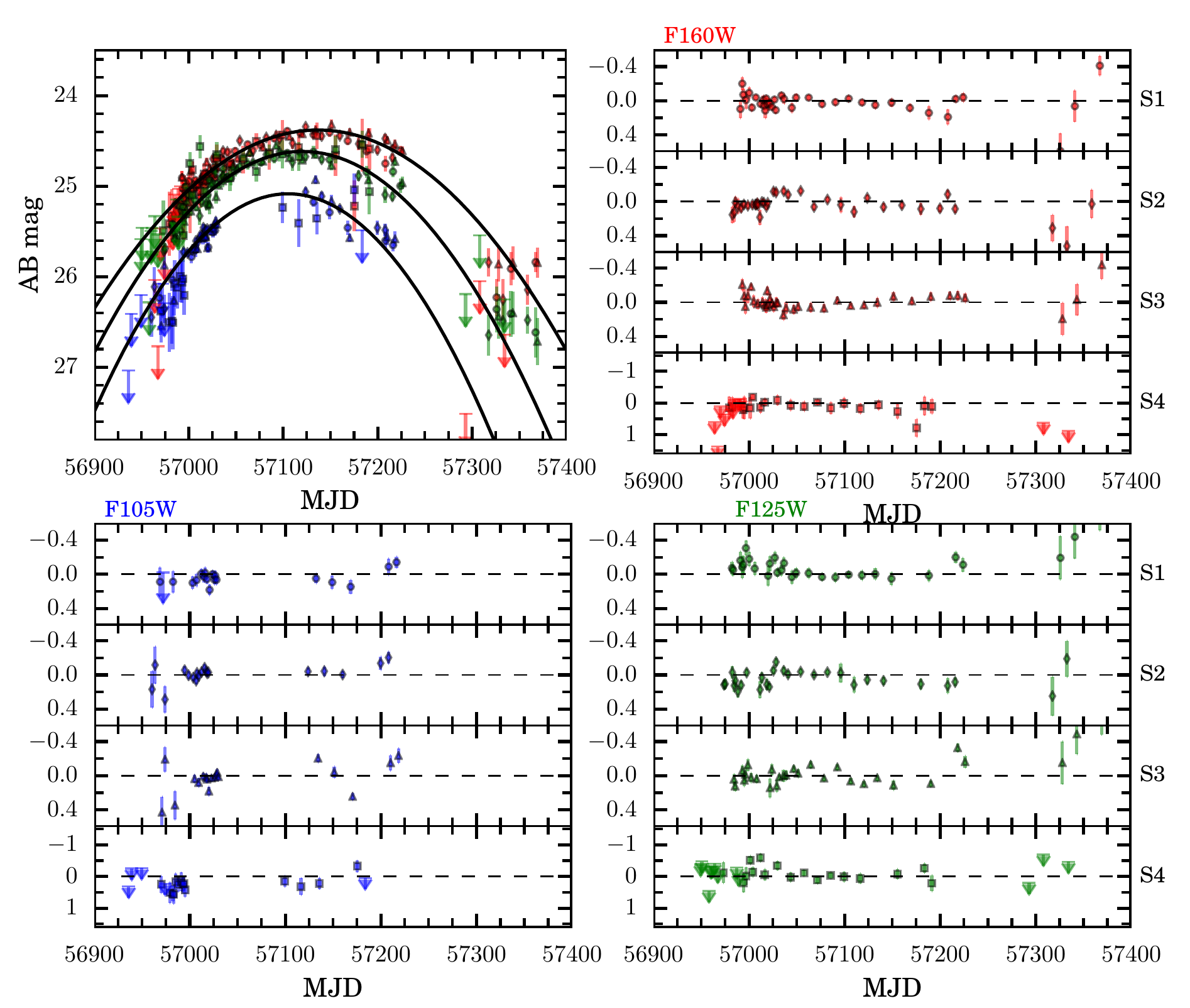}
   \caption{ Fit of the SN Refsdal light curves with the
       simplest viable functional form, a 2$^{\rm nd}$-order Chebyshev
       polynomial. Points in the top left panel show the composite
       S1--S4 light curve data, after applying magnitude and time
       shifts so that S2, S3 and S4 match the reference light curve,
       S1.  The F160W, F125W, and F105W bands are plotted in red,
       green and blue, respectively.  Overplotted curves show the
       maximum likelihood polynomial fit, where separate polynomial
       coefficients are fit for each band. The adjoining panel sets
       show residuals after subtracting off this best-fit model,
       separated by image and band.
   \label{fig:ChebyshevResiduals}}
\end{figure*}

\citet{Dobler:2006} find that most microlensed SN should be expected
to exhibit fluctuations of $\sim0.2$ mag on short timescales
  (days to weeks) and shifts of $>0.5$ mag on long timescales
  (months). These distortions will significantly limit the precision
that can be achieved in measurement of their time delays.  However,
the microlensing environments modeled by \citet{Dobler:2006} are for a
SN lensed by a single isolated galaxy, and the situation for SN
Refsdal may be less dire, as the added shear and convergence
from the \macs1149 cluster potential may result in light paths through
the lensing galaxy that are farther from the galactic nucleus and
therefore intersect relatively sparse stellar environments, even
accounting for the intracluster light.

The SN Refsdal light curve is not as finely sampled or covering as
long a baseline as is now typical for multiply-imaged quasars, which
in some cases are monitored over 5-10 years
\citep[e.g.][]{Courbin:2011,Eulaers:2013,Tewes:2013b}. However, a
microlensing analysis of SN Refsdal would have several distinct
advantages relative to the quasar observations.  First, the intrinsic
SN light curve with a single broad peak is much simpler than the
stochastically varying quasar light curves.  This means that
deviations from a smoothly varying light curve can in principle be
immediately attributed to microlensing variations -- assuming that
systematic uncertainties in the photometry are well controlled.  The
data set for SN Refsdal also includes valuable color information from
simultaneous observations in multiple pass-bands, which has not been
commonly collected for many long-baseline quasar light curves.
Gravitational lensing effects are achromatic in general, so a
  microlensing event should always be independently detected in
  multiple bands.  A limited chromatic dependence can be generated for
  microlensing if the effective source size depends on wavelength
  \citep{Kochanek:2004}, but this will generally be negligible for the
  small wavelength difference between optical and infrared bands
  typically used for SN observations.  Finally, a microlensing
analysis of SN Refsdal will soon be able to take advantage of the
light curve of the fifth image (SX), which is lensed only by the
\macs1149 cluster potential and not also directly affected by any
cluster member galaxies.  The light curve of SX, though expected to be
substantially fainter than S1-S4, should be relatively free of
microlensing, since its light path does not pass through any
individual cluster galaxy, and it should be less affected by
intracluster stars or the outskirts of the \macs1149 Brightest Cluster
Galaxy (BCG).

A full analysis of possible microlensing in the SN Refsdal
  light curves is beyond the scope of this work.  However, we can make
  a preliminary assessment of whether there are any indications of
  especially strong microlensing events that could bias our time delay
  and magnification measurements.  Figure~\ref{fig:ChebyshevResiduals}
  shows the maximum likelihood model derived from fitting three
  2$^{nd}$-order Chebyshev polynomials to the SN Refsdal light curves
  in the F160W, F125W, and F105W bands. This model does not have the
  largest posterior probability among the flexible function models we
  have investigated.  However, it is useful for this analysis because
  it has very limited flexibility, so it cannot be distorted to
  accommodate microlensing events at the edges of the observed data as
  part of the intrinsic light curve shape. Even for this very simple
  model, the residuals shown in Figure~\ref{fig:ChebyshevResiduals}
  are consistently within $\pm$0.2 mag for all epochs between
  MJD$\sim$56950 to 57250.  The final epochs deviate more strongly,
  but this is likely due to the inability of this simple model to
  accommodate a change in slope after the peak.  Although this is far
  from a complete analysis, the absence of significant deviations
  ($>0.2$ mag) relative to this minimally flexible model gives a
  preliminary indication that major microlensing events did not affect
  the SN Refsdal light curves.

\subsection{Future Measurements of SN Time Delays}\label{sec:Future}

In the near future, we may expect that additional detections of
strongly lensed SNe with measurable time delays will be few and far
between.  For most massive galaxy clusters, the rate of such SNe
visible above a magnitude limit of $m_{AB}\sim27.0$ is expected to be
on the order of a few SNe per century
\citep{Gal-Yam:2002b,Bolton:2003,Sharon:2010,Li:2012}. This means that
roughly 100 clusters must be regularly monitored with deep
imaging in order to have a realistic chance of detecting another event
like SN Refsdal in any given year.  The Reionization Lensing Cluster
Survey (RELICS, HST-GO-14096; PI:D. Coe) is an ongoing \HST\ program
that is a step in that direction, with cadenced IR imaging of 46
strong-lensing galaxy cluster fields. However, each RELICS cluster
target is only monitored over a period of 1-2 months, so RELICS and
any similar cluster surveys will still have only a small chance of
discovering another multiply-imaged SN in the near future.  Such
programs are much more likely to find lensed SNe with significant
magnification but no multiple images
\citep{Sullivan:2000,Goobar:2009}, which can still be useful as
a means for discovering distant SNe \citep{Gunnarsson:2003,Amanullah:2011} or
  for testing cluster lens models
  \citep{Riehm:2011,Patel:2014,Nordin:2014,Rodney:2015a}.

Although we have seen that cluster-lensed SNe like SN Refsdal are
valuable as tools for testing cluster dark matter models, another
strong motivation for measuring SN time delays is for cosmological
constraints.  After accounting for the difference in the gravitational
potential traversed by the light path for each multiple image
\citep{Shapiro:1964}, the time delay can be used to directly constrain
the Hubble constant \citep{Refsdal:1964} and other cosmological
parameters ~\citep{Linder:2004,Coe:2009,Linder:2011}.  Distant quasars
multiply-imaged by foreground galaxies have been used for such time
delay cosmography measurements, providing valuable cosmological
constraints that can complement or validate other methods such as Type
Ia SNe, baryon acoustic oscillations and the cosmic microwave
background (e.g.,
\citealt{Saha:2006,Oguri:2007,Coles:2008,Suyu:2010,Suyu:2013,Suyu:2014};
and see \citealt{Treu:2014} for a recent review).

Although rare, cluster-lensed SNe such as SN Refsdal could in
principle contribute to future time delay cosmography
efforts. However, cluster lenses are generally much more complex than
isolated galaxies, which limits the possible precision of time delay
cosmography. A more promising avenue for building up a cosmologically
useful sample of strongly-lensed SNe is through wide-field imaging
surveys that can find lensed SNe behind {\it galaxy-scale} lenses
\citep{Oguri:2003b,Mortsell:2005}.  The first example of this came
from the Pan-STARRS survey \citep{Kaiser:2010}, as SN PS1-10afx
\citep{Chornock:2013} was shown to be strongly-lensed (though not
multiply-imaged) by an intervening galaxy
\citep{Quimby:2013,Quimby:2014}.  The Large Synoptic Survey Telescope
\citep[LSST][]{Tyson:2002} will dramatically extend this sample, as it
is expected to deliver $\sim$130 SNe strongly lensed by foreground
galaxies \citep{Oguri:2010a}.  The Wide Field Infrared Survey
Telescope (WFIRST) could substantially increase that sample,
particularly at the high redshift end
\citep{Holz:2001,Mortsell:2005,Oguri:2010a}.  Time delays for these
galaxy-scale lenses are on the order of days or months, not years or
decades as is typically the case for cluster-scale lenses, which
reduces the timescale over which an observational monitoring campaign
needs to operate.  Furthermore, the lensing potential of a solitary
galaxy -- especially an elliptical -- is much simpler than a typical
galaxy cluster.  This means that with a sufficiently rapid
observational cadence and concerted lens modeling efforts it will be
feasible to use measurements of these lensed SN time delays as
cosmographic tools, which will finally realize the original vision of
SN Refsdal's namesake \citep{Refsdal:1964}.

%% file: TableResultsSummary.tex
\begin{deluxetable}{lcc}
  \tablecaption{\sc Summary of Time Delay and Magnification Ratio Measurements \label{tab:ResultsSummary}}
  \tablehead{
\colhead{} & \colhead{Template}  & \colhead{Polynomial} \\
\colhead{Parameter} & \colhead{LC Fits\tablenotemark{a}}  & \colhead{Curve Fits\tablenotemark{a}} \\
  }
  \startdata    
MJD$_{\rm pk}$            & 57138$\pm$10 days &  57132$\pm$4 days\\
$\Delta t_{S2:S1}$        &  4$\pm$4 days &  7$\pm$2 days \\
$\Delta t_{S3:S1}$        &  2$\pm$5 days &  0.6$\pm$3 days \\
$\Delta t_{S4:S1}$        &  24$\pm$7 days &  27$\pm$8 days \\
$\mu_{\rm S2}/\mu_{\rm S1}$ & 1.15$\pm$0.05 & 1.17$\pm$0.02\\
$\mu_{\rm S3}/\mu_{\rm S1}$ & 1.01$\pm$0.04 & 1.00$\pm$0.01\\
$\mu_{\rm S4}/\mu_{\rm S1}$ & 0.34$\pm$0.02 & 0.38$\pm$0.02\\
  \enddata
\tablenotetext{a}{Mean values from the BMA combinations given in Tables~\ref{tab:TemplateFitResults} and \ref{tab:FlexFitResults}, with uncertainties updated to incorporate the mock light curve analysis of Section~\ref{ssec:FlexFitUncertaintyEstimates}.}
\end{deluxetable}

%% file: Acknowledgments.tex
\begin{acknowledgments}
The authors thank Stefano Casertano, Armin Rest, Piero Rosati, and
Claudio Grillo for helpful discussion of this paper.  

Financial support for this work was provided to S.A.R. by NASA through
grant HST-GO-13386 from the Space Telescope Science Instittute
(STScI), which is operated by Associated Universities for Research in
Astronomy, Inc. (AURA), under NASA contract NAS
5-26555. R.J.F.\ gratefully acknowledges support from NSF grant
AST-1518052 and the Alfred P.\ Sloan Foundation.  This supernova
research at Rutgers University is supported by NSF CAREER award
AST-0847157, and NASA/Keck JPL RSA 1508337 and 1520634 to SWJ.  TT
acknowledges support from NASA through grant HST-GO-13459 for the
GLASS program.  AZ is supported by NASA through Hubble Fellowship
grant \#HST-HF2-51334.001-A awarded by STScI.
\end{acknowledgments}

%% file: LongTablePhotometry.tex
\clearpage
\begin{longtable*}{lrrr@{\,$\pm$\,}lr@{\,$\pm$\,}lr@{\,$\pm$\,}lr@{\,$\pm$\,}l}
  \tablecaption{\sc Photometry of the Four Einstein Cross Images of SN Refsdal\label{tab:Photometry}}
\tablecolumns{11}
\tablehead{
\colhead{Filter} & \colhead{MJD} & \colhead{Exp. Time} & \multicolumn{2}{c}{S1} & \multicolumn{2}{c}{S2} &  \multicolumn{2}{c}{S3} & \multicolumn{2}{c}{S4}\\
\colhead{} & \colhead{} & \colhead{(s)} & \multicolumn{2}{c}{(AB mag)} & \multicolumn{2}{c}{(AB mag)} &  \multicolumn{2}{c}{(AB mag)} & \multicolumn{2}{c}{(AB mag)}}
\multicolumn{11}{p{0.6\textwidth}}{{\sc NOTE--} Magnitudes are reported as 3$\sigma$ lower limits when the measured flux from PSF fitting was less than the flux uncertainty.  Empty entries indicate instances where the PSF fitting did not converge.  Asterisks mark values measured on images processed with the ``despiking'' algorithm illustrated in Figure~\ref{fig:macs1149field}.}\\
\endlastfoot
F160W & 56990.9 &    1159 &    25.22 &    0.11 $^{*}$&    25.20 &    0.10 &    24.91 &    0.07 & \multicolumn{2}{c}{\nodata} \\
F160W & 56993.0 &    1159 &    24.91 &    0.07 $^{*}$&    25.15 &    0.11 &    25.02 &    0.08 & \multicolumn{2}{c}{\nodata} \\
F160W & 56994.0 &    1159 &    25.03 &    0.07 $^{*}$&    25.02 &    0.07 &    25.14 &    0.08 & \multicolumn{2}{c}{\nodata} \\
F160W & 56996.8 &    1159 &    25.06 &    0.10 $^{*}$&    25.04 &    0.09 &    24.99 &    0.08 & \multicolumn{2}{c}{$>$25.96} \\
F160W & 57000.1 &    2318 &    24.95 &    0.07 $^{*}$&    25.02 &    0.06 &    24.85 &    0.04 & \multicolumn{2}{c}{$>$26.23} \\
F160W & 57003.0 &    5512 &    25.09 &    0.04 &    24.95 &    0.04 &    24.99 &    0.07 &    26.85 &    0.40 \\
F160W & 57007.1 &    5512 &    24.93 &    0.03 &    24.93 &    0.05 &    24.98 &    0.06 &    27.05 &    0.56 \\
F160W & 57012.0 &    5512 &    24.96 &    0.05 &    24.87 &    0.04 &    24.92 &    0.06 &    26.44 &    0.36 \\
F160W & 57015.0 &    5512 &    24.89 &    0.05 &    24.84 &    0.04 &    24.93 &    0.07 &    26.35 &    0.30 \\
F160W & 57015.9 &    5512 &    24.95 &    0.04 &    24.83 &    0.05 &    24.95 &    0.06 &    26.72 &    0.42 \\
F160W & 57017.0 &    1015 &    25.00 &    0.09 &    24.81 &    0.05 &    24.75 &    0.05 & \multicolumn{2}{c}{\nodata} \\
F160W & 57017.9 &    5512 &    24.85 &    0.04 &    24.82 &    0.04 &    24.90 &    0.06 &    26.48 &    0.39 \\
F160W & 57019.7 &    1421 &    24.90 &    0.07 &    24.95 &    0.09 &    24.82 &    0.06 & \multicolumn{2}{c}{\nodata} \\
F160W & 57021.9 &    6121 &    24.88 &    0.05 &    24.75 &    0.03 &    24.85 &    0.06 &    26.41 &    0.38 \\
F160W & 57023.9 &    5512 &    24.90 &    0.03 &    24.74 &    0.04 &    24.88 &    0.06 &    26.38 &    0.36 \\
F160W & 57026.6 &   11023 &    24.79 &    0.03 &    24.76 &    0.04 &    24.80 &    0.06 &    26.34 &    0.26 \\
F160W & 57027.9 &    5512 &    24.91 &    0.04 &    24.71 &    0.03 &    24.80 &    0.05 &    26.26 &    0.36 \\
F160W & 57034.0 &    1159 &    24.69 &    0.04 &    24.53 &    0.04 &    24.90 &    0.06 &    26.20 &    0.34 \\
F160W & 57036.6 &    1159 &    24.71 &    0.06 &    24.51 &    0.05 &    24.80 &    0.07 &    25.84 &    0.09 \\
F160W & 57044.7 &    1209 &    24.76 &    0.05 &    24.44 &    0.04 &    24.76 &    0.05 &    26.08 &    0.18 \\
F160W & 57049.2 &    1209 &    24.61 &    0.04 &    24.45 &    0.04 &    24.70 &    0.05 &    25.89 &    0.12 \\
F160W & 57062.4 &    1209 &    24.54 &    0.03 &    24.33 &    0.03 &    24.63 &    0.06 &    25.72 &    0.16 \\
F160W & 57076.4 &    1209 &    24.55 &    0.04 &    24.44 &    0.04 &    24.58 &    0.05 &    25.78 &    0.17 \\
F160W & 57090.4 &    1209 &    24.47 &    0.04 &    24.29 &    0.03 &    24.44 &    0.04 &    25.73 &    0.15 \\
F160W & 57104.3 &    1209 &    24.39 &    0.04 &    24.31 &    0.09 &    24.46 &    0.04 &    25.53 &    0.09 \\
F160W & 57118.2 &    1209 &    24.41 &    0.04 &    24.35 &    0.06 &    24.43 &    0.04 &    25.65 &    0.18 \\
F160W & 57132.1 &     759 &    24.43 &    0.04 &    24.17 &    0.03 &    24.39 &    0.06 &    25.46 &    0.17 \\
F160W & 57149.1 &     759 &    24.41 &    0.05 &    24.27 &    0.05 &    24.33 &    0.05 &    25.59 &    0.16 \\
F160W & 57168.3 &    1209 &    24.50 &    0.05 &    24.27 &    0.04 &    24.42 &    0.05 &    25.46 &    0.15 \\
F160W & 57188.2 &    1209 &    24.61 &    0.08 $^{*}$&    24.36 &    0.07 &    24.43 &    0.05 &    25.68 &    0.23 $^{*}$\\
F160W & 57208.1 &    1209 &    24.75 &    0.09 $^{*}$&    24.43 &    0.04 &    24.51 &    0.05 &    26.24 &    0.28 $^{*}$\\
F160W & 57216.2 &    1209 &    24.58 &    0.04 &    24.30 &    0.05 &    24.55 &    0.05 &    25.57 &    0.29 $^{*}$\\
F160W & 57224.0 &    1209 &    24.61 &    0.05 &    24.51 &    0.05 &    24.62 &    0.05 &    25.61 &    0.23 $^{*}$\\
F160W & 57325.8 &    1259 &    26.23 &    0.21 &    25.67 &    0.16 &    25.87 &    0.19 & \multicolumn{2}{c}{$>$25.84} \\
F160W & 57341.0 &    1259 &    25.91 &    0.18 &    26.08 &    0.23 &    25.86 &    0.18 &    27.43 &    1.08 \\
F160W & 57367.1 &    1259 &    25.84 &    0.11 &    25.97 &    0.17 &    25.86 &    0.17 &    28.02 &    1.50 \\[2mm]
F140W & 56972.1 &    1168 &    25.43 &    0.11 &    25.52 &    0.09 &    25.56 &    0.10 &    26.74 &    0.36 \\
F140W & 56982.4 &   15935 &    25.41 &    0.07 &    25.38 &    0.05 &    25.37 &    0.05 &    27.13 &    0.39 \\
F140W & 56983.2 &    5212 &    25.37 &    0.07 &    25.29 &    0.05 &    25.35 &    0.06 &    27.02 &    0.50 \\
F140W & 56984.9 &    5212 &    25.33 &    0.06 &    25.35 &    0.05 &    25.36 &    0.05 &    27.01 &    0.52 \\[2mm]
F125W & 56982.2 &   10423 &    25.42 &    0.08 &    25.54 &    0.05 &    25.52 &    0.06 &    26.97 &    0.51 \\
F125W & 56983.4 &   10423 &    25.43 &    0.07 &    25.52 &    0.06 &    25.58 &    0.06 &    26.84 &    0.45 \\
F125W & 56990.9 &    1159 &    25.22 &    0.10 &    25.28 &    0.07 &    25.31 &    0.06 &    27.67 &    1.07 $^{*}$\\
F125W & 56993.0 &    1159 &    25.28 &    0.13 &    25.41 &    0.08 &    25.40 &    0.07 &    26.88 &    0.58 $^{*}$\\
F125W & 56994.0 &    1159 &    25.23 &    0.10 &    25.34 &    0.08 &    25.32 &    0.08 &    26.85 &    0.61 $^{*}$\\
F125W & 56996.7 &    1159 &    25.01 &    0.09 &    25.44 &    0.07 &    25.18 &    0.07 &    26.71 &    0.46 $^{*}$\\
F125W & 57000.1 &    2318 &    25.10 &    0.08 &    25.32 &    0.05 &    25.29 &    0.06 &    26.92 &    0.42 $^{*}$\\
F125W & 57005.9 &    5212 &    25.15 &    0.08 &    25.11 &    0.04 &    25.24 &    0.05 &    26.51 &    0.33 \\
F125W & 57020.0 &     406 &    25.10 &    0.12 &    25.16 &    0.10 &    25.21 &    0.11 &    26.53 &    0.56 $^{*}$\\
F125W & 57021.8 &     812 &    24.94 &    0.09 &    25.00 &    0.06 &    24.98 &    0.07 &    26.72 &    0.57 $^{*}$\\
F125W & 57026.9 &    1623 &    24.82 &    0.07 &    25.03 &    0.07 &    25.13 &    0.07 &    26.57 &    0.29 \\
F125W & 57029.6 &    5212 &    24.97 &    0.05 &    25.03 &    0.04 &    25.00 &    0.03 &    26.32 &    0.27 \\
F125W & 57033.9 &    1159 &    24.91 &    0.06 &    24.80 &    0.04 &    24.95 &    0.07 &    25.77 &    0.13 \\
F125W & 57036.6 &    1159 &    24.81 &    0.07 &    24.68 &    0.05 &    24.93 &    0.05 &    26.12 &    0.15 \\
F125W & 57044.7 &    1209 &    24.91 &    0.08 &    24.72 &    0.05 &    24.79 &    0.04 &    25.59 &    0.13 \\
F125W & 57049.2 &    1209 &    24.83 &    0.06 &    24.73 &    0.04 &    24.82 &    0.05 &    26.07 &    0.17 \\
F125W & 57062.4 &    1209 &    24.76 &    0.06 &    24.61 &    0.03 &    24.64 &    0.04 &    25.67 &    0.14 \\
F125W & 57076.4 &    1209 &    24.74 &    0.04 &    24.57 &    0.03 &    24.74 &    0.04 &    25.93 &    0.15 \\
F125W & 57090.4 &    1209 &    24.69 &    0.05 &    24.48 &    0.03 &    24.56 &    0.04 &    25.71 &    0.10 \\
F125W & 57104.3 &    1209 &    24.63 &    0.04 &    24.45 &    0.11 &    24.70 &    0.04 &    25.86 &    0.11 \\
F125W & 57118.2 &    1209 &    24.63 &    0.06 &    24.56 &    0.09 &    24.72 &    0.04 &    25.67 &    0.10 \\
F125W & 57132.1 &     759 &    24.62 &    0.07 &    24.50 &    0.05 &    24.66 &    0.04 &    25.67 &    0.16 \\
F125W & 57149.1 &     759 &    24.71 &    0.07 &    24.53 &    0.05 &    24.78 &    0.05 &    25.70 &    0.16 \\
F125W & 57188.2 &    1209 &    24.84 &    0.07 &    24.71 &    0.06 &    24.94 &    0.04 &    25.62 &    0.15 $^{*}$\\
F125W & 57216.2 &    1209 &    24.84 &    0.06 &    24.92 &    0.09 $^{*}$&    24.74 &    0.05 $^{*}$&    25.56 &    0.15 $^{*}$\\
F125W & 57224.0 &    1209 &    25.00 &    0.08 &    24.94 &    0.05 $^{*}$&    24.97 &    0.07 $^{*}$&    26.08 &    0.25 \\
F125W & 57325.8 &    1159 &    26.36 &    0.25 &    26.47 &    0.23 &    26.45 &    0.25 & \multicolumn{2}{c}{$>$26.22} \\
F125W & 57340.9 &    1159 &    26.41 &    0.25 &    26.32 &    0.21 &    26.41 &    0.23 &    26.92 &    0.55 \\
F125W & 57367.1 &    1159 &    26.62 &    0.28 &    26.31 &    0.15 &    26.73 &    0.26 &    27.63 &    0.91 \\[2mm]
F105W & 56968.9 &    1218 &    26.24 &    0.17 &    26.28 &    0.21 &    26.55 &    0.18 & \multicolumn{2}{c}{$>$26.60} \\
F105W & 56972.1 &     356 &    26.44 &    0.53 &    25.94 &    0.22 &    25.89 &    0.15 & \multicolumn{2}{c}{$>$26.18} \\
F105W & 56982.4 &    5612 &    26.03 &    0.13 &    26.18 &    0.16 &    26.27 &    0.17 &    27.58 &    0.65 \\
F105W & 57003.0 &    5612 &    25.78 &    0.07 &    25.56 &    0.04 &    25.70 &    0.05 &    27.40 &    0.27 \\
F105W & 57007.1 &    5612 &    25.70 &    0.08 &    25.57 &    0.05 &    25.70 &    0.05 &    27.68 &    0.46 \\
F105W & 57012.0 &    5612 &    25.58 &    0.06 &    25.53 &    0.03 &    25.58 &    0.04 &    27.52 &    0.33 \\
F105W & 57015.0 &    5612 &    25.58 &    0.06 &    25.53 &    0.04 &    25.56 &    0.05 &    27.51 &    0.25 \\
F105W & 57015.9 &    5612 &    25.52 &    0.06 &    25.43 &    0.04 &    25.57 &    0.04 &    27.53 &    0.29 \\
F105W & 57017.9 &    5612 &    25.58 &    0.05 &    25.45 &    0.04 &    25.69 &    0.05 &    27.10 &    0.26 \\
F105W & 57020.7 &    5612 &    25.68 &    0.06 &    25.37 &    0.04 &    25.51 &    0.04 &    27.14 &    0.23 \\
F105W & 57023.9 &    5612 &    25.46 &    0.04 &    25.29 &    0.04 &    25.47 &    0.04 &    27.02 &    0.29 \\
F105W & 57025.7 &    5612 &    25.51 &    0.06 &    25.31 &    0.03 &    25.43 &    0.03 &    27.08 &    0.23 \\
F105W & 57026.6 &    5612 &    25.46 &    0.06 &    25.32 &    0.03 &    25.39 &    0.04 &    27.04 &    0.21 \\
F105W & 57027.9 &    5612 &    25.49 &    0.06 &    25.30 &    0.03 &    25.43 &    0.04 &    27.23 &    0.22 \\
F105W & 57132.1 &     759 &    25.18 &    0.05 &    24.89 &    0.05 &    24.94 &    0.05 &    26.26 &    0.17 \\
F105W & 57149.1 &     759 &    25.29 &    0.08 &    24.94 &    0.05 &    25.17 &    0.07 &    26.43 &    0.27 \\
F105W & 57168.3 &    1209 &    25.46 &    0.08 &    25.08 &    0.04 &    25.58 &    0.04 &    26.38 &    0.17 \\
F105W & 57208.1 &    1209 &    25.60 &    0.10 &    25.29 &    0.08 &    25.57 &    0.09 &    26.06 &    0.18 $^{*}$\\
F105W & 57216.3 &    2412 &    25.65 &    0.07 &    25.31 &    0.07 &    25.59 &    0.08 $^{*}$&    26.87 &    0.38 $^{*}$\\[2mm]
F814W & 56985.9 &    4768 &    27.98 &    0.45 &    27.99 &    0.56 & \multicolumn{2}{c}{$>$27.43} & \multicolumn{2}{c}{\nodata} \\
F814W & 57014.7 &   10616 &    27.73 &    0.45 &    27.23 &    0.27 &    27.30 &    0.29 & \multicolumn{2}{c}{$>$27.58} \\
F814W & 57132.6 &    9826 &    26.88 &    0.13 &    26.90 &    0.10 &    26.96 &    0.17 &    28.15 &    0.52 \\
F814W & 57135.6 &   14538 &    27.16 &    0.12 &    26.82 &    0.07 &    27.04 &    0.14 &    27.59 &    0.26 \\
F814W & 57138.2 &   19652 &    27.11 &    0.14 &    26.82 &    0.08 &    27.25 &    0.14 &    28.06 &    0.29 \\
F814W & 57143.8 &   29880 &    27.20 &    0.12 &    26.98 &    0.09 &    27.22 &    0.17 &    28.32 &    0.43 \\
F814W & 57151.6 &   14943 &    27.35 &    0.21 &    27.04 &    0.20 &    27.18 &    0.18 &    27.82 &    0.27 \\
F814W & 57159.9 &   10092 &    27.48 &    0.21 &    27.18 &    0.19 &    27.30 &    0.20 &    27.77 &    0.26 \\[2mm]
F606W & 56985.8 &    4542 & \multicolumn{2}{c}{$>$27.89} &    28.37 &    0.47 & \multicolumn{2}{c}{$>$27.75} & \multicolumn{2}{c}{\nodata} \\
F606W & 57151.6 &   14724 & \multicolumn{2}{c}{$>$27.99} & \multicolumn{2}{c}{$>$27.74} & \multicolumn{2}{c}{$>$28.00} &    29.14 &    0.80 \\
F606W & 57161.4 &   10092 & \multicolumn{2}{c}{$>$27.89} & \multicolumn{2}{c}{$>$27.90} & \multicolumn{2}{c}{$>$27.65} & \multicolumn{2}{c}{$>$28.07} \\[2mm]
F435W & 57131.7 &    4744 &    27.67 &    0.52 & \multicolumn{2}{c}{$>$27.53} & \multicolumn{2}{c}{$>$27.71} & \multicolumn{2}{c}{$>$27.61} \\
F435W & 57138.2 &    9488 &    27.88 &    0.48 & \multicolumn{2}{c}{$>$27.51} & \multicolumn{2}{c}{$>$27.64} & \multicolumn{2}{c}{$>$27.63} \\
\hline\\[2mm]
\end{longtable*}
\clearpage